\documentclass[12pt]{article}

\usepackage{amssymb,graphicx}
\usepackage{epsf}
\usepackage{graphicx,epsfig}
\usepackage{amsfonts}
\usepackage{amssymb}
\usepackage{color}

\def\bk{{\bf k}}

\def\CO{{\cal O}}

\def\high{\vphantom{\Biggl(}\displaystyle}

\def\mpl{M_{\rm Pl}}
\def\half{\frac{1}{2}}

\makeatletter
\renewcommand\section{\@startsection {section}{1}{\z@}%
                                 {-3.5ex \@plus -1ex \@minus -.2ex}
                                   {2.3ex \@plus.2ex}%
                                   {\normalfont\large\bfseries}}
\renewcommand\subsection{\@startsection{subsection}{2}{\z@}%
                                   {-3.25ex\@plus -1ex \@minus -.2ex}%
                                     {1.5ex \@plus .2ex}%
                                     {\normalfont\bfseries}}
\renewcommand\subsubsection{\@startsection{subsubsection}{3}{\z@}%
                                   {-3.25ex\@plus -1ex \@minus -.2ex}%
                                     {1.5ex \@plus .2ex}%
                                     {\normalfont\itshape}}
\makeatother

\newcommand{\Letter}{
\setlength{\textwidth}{16.5cm}
   \setlength{\textheight}{22.6cm}
    \hoffset=-0.5in
\voffset=-2.1cm }

\Letter

\begin{document}
\newcommand{\be}{\begin{equation}}
\newcommand{\ee}{\end{equation}}
\newcommand{\bea}{\begin{eqnarray}}
\newcommand{\eea}{\end{eqnarray}}
\newcommand{\barr}{\begin{array}}
\newcommand{\earr}{\end{array}}
\newcommand{\myfigure}[2]{\resizebox{#1}{!}{\includegraphics{#2}}}

\thispagestyle{empty}
\begin{flushright}
\parbox[t]{1.6in}{MIT-CTP-3870}
\end{flushright}

\vspace*{0.7in}

\begin{center}
{\large \bf Comparing Infrared Dirac-Born-Infeld Brane Inflation to Observations}

\vspace*{0.4in} {Rachel Bean$^1$, Xingang Chen$^{2,3}$, 
Hiranya Peiris$^{4,5,}$\footnotemark[0]\footnotetext{Hubble Fellow} and Jiajun Xu$^3$}
\\[.3in]
{\em $^1$ Department of Astronomy, Cornell University, Ithaca, NY
14853, USA \\
$^2$ Center for Theoretical Physics, \\
Massachusetts Institute of Technology, Cambridge, MA 02139, USA \\
$^3$ Newman Laboratory for Elementary Particle Physics,\\ Cornell
University, Ithaca, NY 14853, USA \\
$^4$ Kavli Institute for Cosmological Physics and Enrico Fermi Institute,\\
 University of Chicago, Chicago IL 60637, USA \\
 $^5$ Institute of Astronomy, University of Cambridge, Cambridge CB3 0HA, UK}
\\[0.3in]
\end{center}

\begin{center}
{\bf
Abstract}
\end{center}
\noindent
We compare the Infrared Dirac-Born-Infeld (IR DBI) brane inflation
model to observations using
a Bayesian analysis. The current data cannot distinguish
it from the {\rm 
$\Lambda$CDM} model, but is able to give interesting constraints on
various microscopic parameters including the mass of the brane moduli
potential, the fundamental string scale, the charge or warp factor of
throats, and the number of the mobile branes. We quantify some
distinctive testable predictions 
with stringy signatures, such as the large non-Gaussianity, and the
large, but regional, running of the spectral index. These results
illustrate how we may be able to probe aspects of string theory
using cosmological observations. 

\vfill

\newpage
\setcounter{page}{1}

\tableofcontents

\newpage

\section{Introduction}\label{Sec:intro}
\setcounter{equation}{0}

\subsection{Experiments and theories}
\label{Sec:ExTh}

An ongoing and forthcoming array of experiments (e.g. WMAP,
\cite{Spergel:2006hy}, SDSS \cite{York:2000gk, Tegmark:2003uf, Tegmark:2006az, McDonald:2004xn}, SNLS
\cite{Astier:2005qq}, ACBAR \cite{Kuo:2006ya}, Planck \cite{Planck},
ACT \cite{ACT}, Spider \cite{Spider}) is measuring the cosmic
microwave background (CMB) and the large scale structure of the
universe with unprecedented precision. This provides exciting
opportunities to reveal the nature of the early universe and the
underlying fundamental theories. The leading theoretical candidate for
creating the initial conditions of our universe is inflation
\cite{Guth:1980zm,Linde:1981mu,Albrecht:1982wi}. However inflation
remains a paradigm, which can be implemented by a variety of models
underpinned by differing microphysical constructions; as the
constraints from data tighten, there is the hope that we might
identify the specific scenario that describes our universe. With the
natural ingredients of such model-building being supergravity and
string theory, the process of better measuring the properties of the
early universe is also a process of understanding better the theory of
quantum gravity.

In contrast to recent debates on the predictivity of the string theory
landscape, here we use a more conventional approach to investigate the
predictivity of string theory by studying the properties and exploring
the dynamics of our own vacuum. We first scan the parameter space of
inflationary models subject only to the requirement that they provide
enough inflationary $e$-folds to solve the flatness and horizon
problems. This is because the natural creation of a homogeneous and
isotropic universe is the leading problem that we want to solve, and
is perhaps the most attractive feature of the inflationary
paradigm. After that, we study the observational consequences of all
the viable parameter spaces, with the goal of looking for distinctive
signatures. Some of these can be compared with observations and used
to narrow down the parameter space. Despite of the vastness of all
possible vacua in the string landscape, this process can be rather
effective since certain observational features rely on distinctive
dynamics.
As we will see,
such dynamics can either be field-theoretic with strong motivations
from string theory, or completely stringy in nature. 

There are many candidate observable signatures in inflationary
models. The most generic ones are the amplitude of the primordial
power spectrum and its spectral index. Since most viable models built
from a fundamental theory have adjustable parameters to fit these two
observables, this leaves a large number of viable models that are
consistent with the data, and even leaves the nature of the inflaton
field ambiguous. In principle, Nature is not obligated to provide more
information within our experimental abilities, and indeed there is no
evidence for further parameters required to describe the current
data. But anticipating her generosity, possible distinctive
observables that might be measurable in the future include the
scale-dependence  (``running'') of the spectral index, departures from
Gaussianity of the primordial fluctuations, a tensor contribution to
the primordial power spectrum, and cosmic strings. These will be
crucial to successfully carry out the program that we have outlined.

With the rapidly improving quality of cosmological data, it will
become increasingly interesting to implement the above program by
comparing specific models to data, starting directly from microscopic
parameters of theories. Modern cosmological data analyses make use of
the powerful method known as Markov Chain Monte Carlo (MCMC) to
implement the comparison to data, providing an efficient way of
estimating posterior distributions of the microscopic
parameters. However, in practice, when directly using microscopic
parameters as MCMC parameters, highly non-linear relationships between
the parameters and observables may introduce severe obstacles for MCMC
to efficiently search the parameter space. Therefore, a
reparameterization according to the specific nature of the model often
becomes necessary. So instead of a straightforward exercise,
implementing MCMC becomes a rather interesting model-dependent art. It
is also a purpose of this paper to use an example to illustrate this
process and extract certain model-independent procedures of 
such reparameterization which may be of more general interest.

\subsection{Brane inflation}

The inflationary models that we study in this paper belong to the
brane inflation scenario proposed by Dvali and Tye
\cite{Dvali:1998pa,HenryTye:2006uv}.\footnote{For recent reviews on
other types of string inflation models, see
Ref.~\cite{Burgess:2007pz,Linde:2007fr,Kallosh:2007ig,Cline:2006hu}.}
We are interested in these models precisely because they can give rise
to a large number of distinctive observational signatures. This
happens even in the simplest scenarios that provide inflation. One of
the most important reasons that makes it possible is that brane
inflation can be achieved via two different mechanisms, namely
slow-roll and Dirac-Born-Infeld (DBI) inflation.

The original models of brane inflation
\cite{Dvali:1998pa,Burgess:2001fx,Dvali:2001fw,Kachru:2003sx} are
slow-roll inflationary models \cite{Linde:1981mu,Albrecht:1982wi},
where branes and anti-branes slowly approach each other in a flat
potential. A model that uses this mechanism in the framework of the
string theory flux compactification \cite{Douglas:2006es} is studied
by Kachru, Kallosh, Linde, Maldacena, McAllister and
Trivedi (KKLMMT) \cite{Kachru:2003sx}. As in the F-term inflation
models in supergravity \cite{Lyth:1998xn},
it is found that the generic shape of the potential is too steep to
achieve the slow-roll inflation, in this case due to the moduli
stabilization. Again, similar to those supergravity models, it is
possible that several contributions to the potential manage to cancel
to a certain precision so that the potential becomes sufficiently
flat. There are effective parameters in the model controlling the
inflaton mass that can be adjusted to fit the observed spectral index
\cite{Bean:2007hc}. The running of the spectral index,
non-Gaussianities and tensor modes are all too small to be observed in
the near future.\footnote{Some observables become measurable if there
are sharp features in the potential
\cite{Adams:2001vc,Peiris:2003ff,Covi:2006ci,Chen:2006xj,Hailu:2006uj}.
In addition, there are other important observational possibilities of
brane inflation -- cosmic 
strings and those related to reheating
\cite{Polchinski:2004ia,HenryTye:2006uv} --
which apply to both slow-roll and DBI inflation.}

Another inflationary mechanism that is so far uniquely found in brane
inflation is the DBI inflation
\cite{Silverstein:2003hf,Alishahiha:2004eh,Chen:2004gc,Chen:2005ad}.
In DBI inflation, the rolling velocity of inflaton branes is not
determined by the shape of the potential but by the speed-limit of the
warped internal space. Such warped spaces are naturally present in the
extra dimensions due to fluxes used to stabilize the string
compactification \cite{Giddings:2001yu}.

The first model that uses such a mechanism is that of Silverstein, Tong and Alishahiha (STA) \cite{Silverstein:2003hf,Alishahiha:2004eh}. In this model, as the branes roll into a throat from the UV side of the warped space under a quadratic potential, its velocity gets restricted by the large warping in the IR side of the warped space. However, instead of having a potential with a generic mass term, a
rather steep potential, characterized by a large inflaton mass, is
required to achieve this UV DBI inflation. The reason is that, when
the branes enter from the UV side of the warped space in the GKP-type
warped compactification \cite{Giddings:2001yu}, the energy
provided by the antibranes sitting at the IR side is not large enough
to drive DBI inflation even if there is the speed-limit, since the
antibrane tension has been warped down correspondingly. Therefore an
extra, steep, potential has to be added to raise the inflationary
energy. In addition, embedded in the same warped compactification,
the model generates large non-Gaussianities that
exceed the experimental bound
\cite{Bean:2007hc,Peiris:2007gz}, as well as excessive probe brane
backreaction which we will address in Sec.~\ref{Sec:UV}. This is because
in this model, the levels of non-Gaussianity and probe brane backreaction 
sensitively depend on the
inflaton value, and it is viable only if the inflaton field is of
(super-)Planckian size. However, the range of the inflaton field is
restricted by some geometric conditions of the compactification and is
sub-Planckian
\cite{Chen:2005fe,Chen:2006hs,Baumann:2006cd,Bean:2007hc,Peiris:2007gz}. 

To fully make use of the speed-limit of the warped space, it is better to make the branes roll out from the IR end,
and use antibranes in other throats to provide the inflationary energy. In this way the speed-limit of the branes 
and the inflationary energy become relatively independent of each
other, leaving a rather flexible shape of the inflaton potential which
has been the main problem of model-building. This is the model
proposed in Ref.~\cite{Chen:2004gc,Chen:2005ad}. It can be generically
realized in the multi-throat brane inflation scenario
\cite{Chen:2004gc}. 

It happens that in this IR DBI inflation model, the large
non-Gaussianities can also be small enough to satisfy the current
observational bound \cite{Chen:2005fe}. This is partly because no
matter how small the warp-factor (and consequently, how big the
non-Gaussianity) the branes begin with, the level of non-Gaussianity
decreases as the branes roll out and approaches its minimal value at
the end of the inflation. Therefore, in the segment of the warped
space traversed during the last 60 $e$-folds, the level of
non-Gaussianity is among the smallest in the entire DBI inflation
trajectory. Moreover, the geometric conditions that put a tight
constraint on the STA model are automatically satisfied in the IR DBI
model and has no effect on the non-Gaussianities.

Besides providing a speed-limit to the inflaton, another important
property of warped space is the reduction of the local fundamental
string scale \cite{Randall:1999ee}. This turns out to have important
consequences on density perturbations in DBI inflationary
models. During the epoch when the string scale is red-shifted below
the Hubble parameter, the quantum fluctuations on the inflaton branes
become stringy.\footnote{Notice that such a stringy phase only happens
in the inflaton sector, which is the deep IR side of a warped space
with energy density of order $H^4$, so it does not backreact
significantly on the Hubble expansion. We also note that such a
stringy phase will backreact on the IR side of the warped geometry,
but it is estimated that this still leaves a large enough portion of
the geometry for DBI inflation to take place
\cite{Chen:2005ad,Chen:2006ni}. We will discuss this more in
Sec.~\ref{Sec:IR} \& \ref{Sec:Power}.}
The density perturbations are no longer fully described by the usual
field theory approximation, and acquire distinctive stringy
signatures. In the IR DBI model, this stringy phase
corresponds to earlier inflationary $e$-folds, and therefore larger
scales in the sky. It is estimated that such a phase transition will
give rise to a large transient (regional) running of the spectral
index \cite{Chen:2005ad,Chen:2005fe}. In this paper, we make this
prediction more quantitative and compare it to observations.

\subsection{Outline}

Following the strategy that we outlined in Sec.~\ref{Sec:ExTh}, in
this paper, we first summarize the overall features of brane inflation
using phase diagrams that describe the parameter spaces spanned by
both inflationary mechanisms, \emph{i.e.} slow-roll vs. DBI
(Sec.~\ref{Sec:phase}), reviewing the key observational predictions in
the different parts of the parameter space. 

The main focus of this paper is to compare the IR DBI brane inflation model to observations (Sec.~\ref{Sec:IRDBI}). We derive analytical and numerical model predictions for the shape of the power spectrum, non-Gaussianity, and tensor modes, giving a quantitative estimate of the effect of the Hubble-expansion-induced stringy phase transition on density perturbations (Appendix \ref{Sec:Width}).

We then proceed to compare these results to the observational data from cosmic microwave background and large scale structure (Sec.~\ref{Sec:MCMC}). We outline how such a comparison should be generally
implemented using MCMC. The current data give a number of interesting
constraints on the microscopic parameters of the model
(Sec.~\ref{conc}), including the mass of the brane moduli potential,
the fundamental string scale, the charge or warp factor of throats, 
and the number
of the mobile branes. We also quantify some distinctive observable
signatures of this model, such as the level of the non-Gaussianity and
the running of the spectral index. We discuss how the latter is
observationally different from two other cases that may also give
large running spectral index: slow-roll inflation with mild features
on potential (Appendix \ref{Sec:Mild}), and
slow-roll or DBI inflation with a non-Bunch-Davies vacuum (Appendix
\ref{Sec:nonBD}). These results illustrate how string theory can make testable predictions which might be subject to observational constraints.

For convenience, all the variables used in this paper are summarized 
in Table~\ref{Tb:variables}.

\begin{table*}[!ht]
\begin{center}
{\small
\begin{tabular}{cll}
\hline
Variable &  Description & Notes\\
\hline \hline
$\mpl$ & 4d reduced Planck mass & 
$\mpl = (8\pi G)^{-1/2}=2.4\times 10^{18} {\rm GeV} $\\
$m_s$ & Mass scale of fundamental strings & $m_s \equiv
\alpha'^{-1/2}$  \\
$g_s$ & String coupling & $g_s<1$ \\
$T_3$ & D3-brane tension &  Eq.~(\ref{RTDef}) \\
$R$ & Length scale of warped throat & Eq.~(\ref{RTDef}) \\
$M$, $K$ & Flux numbers in warped throat & Integers \\
$n_A$ & Number of antibranes in A-throats &  \\
$n_B$ & Number of branes (inflatons) in B-throat & \\
$N_B$ & Effective charge of B-throat & $N_B = a_B MK$ \\
$a_B$ & Multiplicative factor from orbifolding 
& $a_B\sim 1$ in data analysis \\
$\lambda_B$ & $\lambda_B = n_B N_B/2\pi^2$ & Eq.~(\ref{lambdaDef}) \\
$r$ & Radial coordinate of throats &  \\
$\phi$ & Canonical inflaton field & $\phi=r\sqrt{nT_3}$ \\
$h_A$ & Minimum warp factor of A-throat &  \\
$h_B(\phi)$ & Warp factor at location $\phi$ in B-throat & $h\equiv
r/R=\phi R/\sqrt{\lambda}$ \\
$\beta$ & Characterization of shape of potential &
Inflaton mass $m^2 = \beta H^2$ \\
$\gamma$ & Lorentz factor of inflaton & \\
$c_s$ & Sound speed in 4d & $c_s = 1/\gamma$ for DBI inflation \\
$V_0$ & Inflationary energy density & \\
$N_e$ & Number of $e$-folds to the end of inflation\\ 
$N_e^{\rm DBI}$ & Number of $e$-folds to the end of IR DBI inflation& \\
$N_{\rm tot}^{\rm NR}$ & Total $e$-folds of non-relativistic
roll inflation & Typically fast-roll \\
$k_c$ & Critical scale of the stringy phase transition & Eq.~(\ref{kcdef})\\
$N_c$ & Critical DBI $e$-fold at $k_c$ & Eq.~(\ref{Ncdef})\\
$P(k)$ & Power spectrum &\\
$r_{TS}$ & Tensor to scalar ratio & \\
$f_{NL}^{\rm eq}$ & Estimator of the non-Gaussianity & Equilateral shape\\
\hline
\end{tabular}
}
\caption{\label{Tb:variables}
\small Description of variables. In the text, 
subscripts $A$ and $B$ are frequently added to some of the
variables, referring to the quantities of the A- or B-throat.}
\end{center}
\end{table*}

\section{Phase diagrams of brane inflation}\label{Sec:phase}
\setcounter{equation}{0}

As mentioned in the introduction, a useful approach to study
inflationary models is to first scan through as large a parameter
space as possible with the requirement of a sufficient number of
inflationary $e$-folds. Then we can work out the observable
predictions (such as density perturbations) in different regimes, and
compare them with the data to narrow down the parameter space. 

Bearing this in mind, in this section we will study the parameter space in brane inflation models that can provide enough inflationary $e$-folds. In this paper we choose the representative examples in which D3-branes move along the radial direction of a throat with an approximate AdS geometry in type IIB flux compactification.

In the case of the flat 4-d space-time and non-compact extra
dimensions, D3-branes move freely in the throats. However realistic
inflation models in warped compactification requires an inflationary
4-d space-time with a Hubble parameter $H$ and stabilized compact
extra dimensions. In this case the moduli space of branes are lifted
and receive potentials with masses of order $H$. This is only the
generic expectation -- the details of the potential profiles are
environmental, depending on various ingredients (such as fluxes and
other branes) present in specific string compactification models.

In this paper we choose to simply parametrize such unknown mass terms
in the potentials, hoping it can provide a bridge between the bottom-up
observational data-fitting and top-down string theory calculations.

\subsection{UV models}\label{Sec:UV}

In this and the next subsection,
we draw the phase diagrams of brane inflation in
terms of two parameters: the inflaton position $\phi$ and the mass of
the inflaton moduli potential $m$. We use these diagrams to show
the conditions under which different inflationary mechanisms
happen. 

\begin{figure}[t]
\begin{center}
\includegraphics[scale=0.5]{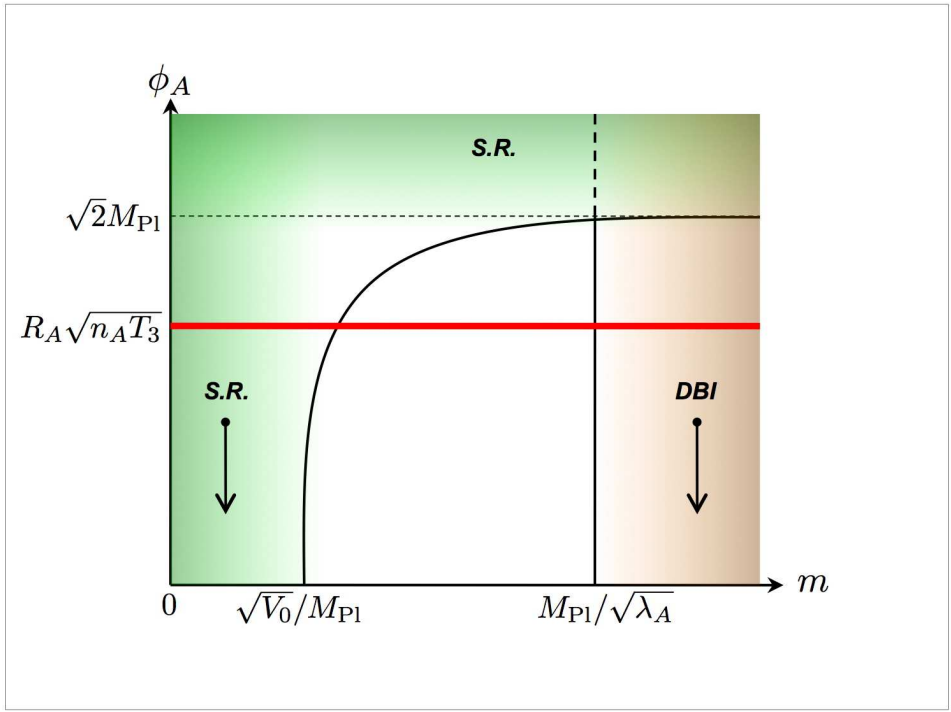}
\end{center}
\caption{\small The inflation phase diagram for UV models. The shaded regions
correspond to parameter space that can give rise to inflation. The
darker the region is, the larger $e$-folds it can provide. ``S.R.''
stands for slow-roll inflation; ``DBI'' stands for DBI inflation. The
arrows indicate the starting point and rolling direction of the
inflaton. In brane inflation, inflatons have to stay below the
horizontal solid line (at $\phi_A = R_A \sqrt{n_A T_3}$); 
the two vertical lines (at $m=\sqrt{V_0}/\mpl$ and
$m=\mpl/\sqrt{\lambda_A}$) are widely separated. 
The curve stretching from
$m=\sqrt{V_0}/\mpl$ to $\phi_A=\sqrt{2}\mpl$ corresponds to
$\eta_V=\beta/3=1$. See text for discussion.}
\label{Fig:UVmodels}
\end{figure}

We first consider the UV models, the phase diagram for which is shown in Fig.~\ref{Fig:UVmodels}. 
In the UV models, the branes are started from the UV side of a
throat (denoted as the A-throat) 
and attracted to the IR end by the moduli potential or Coulomb
potential from antibranes,
\bea
V(\phi) = V_0 + \half m^2 \phi^2 + V_{\rm Coulomb}(\phi) ~,
\label{UVpot}
\eea
where $V_0= 2n_A h_A^4 T_3$ are provided by $n_A$ antibranes at
the end of the
throat, which eventually get annihilated by the same number of 
branes. The warped
geometry is
\bea
ds^2 \propto \frac{\phi^2}{\sqrt{\lambda}} ds_4^2 + 
\frac{\sqrt{\lambda}}{\phi^2} d\phi^2
~,
\label{throatmetric}
\eea
where the $ds_4$ is the 4-d space-time metric, and 
\bea
\lambda_A \equiv n_A
T_3 R_A^4 = n_A N_A/(2\pi^2)
\label{lambdaDef} 
\eea
where $N_A$ and $R_A$ are the effective 
charge and characteristic length scale of the warped space,
respectively.
$n_A$ is the number of inflaton branes.
Note that $N_A$ may include the multiplication factor $a_A$
from orbifolding on
the original D3-charge $N_{0A}$, $N_A \equiv a_A N_{0A}$.
The following relations are also useful:
\bea
R_A^4 = 4 \pi g_s N_A m_s^{-4} ~, ~~~~~
T_3 = \frac{m_s^4}{(2\pi)^3 g_s} ~,
\label{RTDef}
\eea
where $m_s = \alpha'^{-1/2}$ is the string mass scale.
(Later we will also use the same definitions for other throats with
corresponding subscripts $A$ or $B$.)

If the flatness of the potential $V(\phi)$ satisfies the slow-roll
conditions, the branes can slowly roll non-relativistically 
and the kinetic
term of the brane DBI action reduces to the minimal non-relativistic form.
To indicate this condition in the phase diagram, we draw a curve of
$\eta_V \equiv \mpl^2 V''/V =1$ which is
\bea
\phi^2 = 2\mpl^2 - \frac{2V_0}{m^2} ~,
\label{UVslowroll}
\eea
where we neglected the Coulomb potential term for simplicity. 
Eq.~(\ref{UVslowroll}) corresponds to the solid line stretching 
from the lower-left (at $m=\sqrt{V_0}/\mpl$) to
upper-right (at $\phi=\sqrt{2}\mpl$) 
in Fig.~\ref{Fig:UVmodels}. Inclusion of the Coulomb term will cause a slight
deformation at the lower-left corner of this curve and will not
affect our conclusion. The shaded region above and to the left
of this curve has $\eta_V <1$, and corresponds to the slow-roll
inflation phase; here the condition $\epsilon_V<1$ is always weaker.
The lower-left shaded region is the small-field
slow-roll region where the potential is dominated by the constant
$V_0$. As $\phi$ increases toward the upper part of the shaded region, it
corresponds to large-field slow-roll inflation where the potential is
dominated by $m^2 \phi^2$ term. Note that in this region $\phi$ is
trans-Planckian, $\phi>\sqrt{2} \mpl$.

Outside of this slow-roll region, naively the inflaton will roll down
the potential very
fast and make inflation impossible. However because of the presence of
the warped space, the velocity of the inflaton is bounded by the warped
speed of light and therefore cannot be arbitrarily
increased. Silverstein and Tong \cite{Silverstein:2003hf} show that, if 
\bea
m \gg \mpl/\sqrt{\lambda_A} ~, 
\label{UVmcond}
\eea
the inflaton enters another phase of inflation, namely DBI
inflation, in which the full form the DBI kinetic term has to be taken
into account. We indicate this condition by the solid vertical line
(at $m=\mpl/\sqrt{\lambda_A}$) in
Fig.~\ref{Fig:UVmodels} 
and the DBI inflation phase by the shaded region to the right.

There are two possible regions where these two phases merge
onto each other: when the inflaton in the DBI phase starts from a
Planckian value (the upper-right corner of Fig.~\ref{Fig:UVmodels}),
the inflation can
go continuously from slow-roll to DBI; when
$\sqrt{V_0}/\mpl \sim \mpl/\sqrt{\lambda_A}$ so the two vertical lines
in Fig.~\ref{Fig:UVmodels} 
become very close to each other or even switch places, the
inflaton around this border will trigger an inflationary phase that lies
in-between the slow-roll and DBI regimes. These are the ``intermediate
regions'' studied in Refs.~\cite{Shandera:2006ax,Bean:2007hc}.

However, so far we have been discussing an effective field theory
description where one is allowed to independently choose the throat charge $N_A$,
fundamental string mass $m_s$ and the inflaton field range $\phi$. 
In a realistic string
compactification like the generic multi-throat flux compactification
in type IIB string theory \cite{Giddings:2001yu}, 
various throats are glued to a bulk. There are
several fairly model-independent geometric 
conditions that should be imposed to be
consistent with the brane inflation setup that we have in mind.
In this type of compactification, the Planck mass is obtained by an
integration throughout the compact space,
\bea
\mpl^2 \sim g_s^{-2} m_s^8 V_6 ~,
\label{Mplcond}
\eea
where $V_6$ is the total volume of the compactification
(after modding out the possible orbifold effect)
and its dominant contribution
comes from the bulk and the UV regions of throats.
The throats are glued to the bulk 
and their sizes $R$ are restricted,
\bea
\frac{R_A^6}{a_A} \lesssim V_6 ~.
\label{Rcond}
\eea
The volume of the throat is divided by $a_A$ in case of orbifolding.
The inflaton brane separation is restricted by
\bea
\phi \lesssim R_A \sqrt{n_A T_3}~, ~~~
{\rm or}~~~ \phi \lesssim L\sqrt{n_A T_3}
~,
\label{phicond}
\eea
for $n_A$ 
branes moving in the throat or the bulk respectively.
Here $L$ is the size of the bulk in
a certain dimension.
These conditions have been used in various contexts in brane
inflation, to constrain slow-roll models in the bulk
\cite{Burgess:2001fx,Kachru:2003sx}, random walk eternal
inflation \cite{Chen:2006hs},
the UV DBI model through the non-Gaussianities
\cite{Chen:2005fe,Baumann:2006cd,Bean:2007hc,Peiris:2007gz},
and the tensor mode
\cite{Baumann:2006cd,Lidsey:2007gq,Lidsey:2006ia,Spalinski:2007qy}.
For our phase diagram these imply
\bea
\frac{\sqrt{V_0}}{\mpl} \Bigg/ \frac{\mpl}{\sqrt{\lambda_A}} \lesssim
\frac{n_A h_A^2}{N_A/a_A} \ll 1 ~,
\label{UVcond1}
\eea
and
\bea
\frac{R_A\sqrt{n_A T_3}}{\mpl} \lesssim 
\frac{\sqrt{n_A}}{\sqrt{N_A/a_A}} \ll 1 ~.
\label{UVcond2}
\eea
Thus the two vertical solid lines (at $m=\sqrt{V_0}/{\mpl}$ and
$m=\mpl/\sqrt{\lambda_A}$ respectively) in Fig.~\ref{Fig:UVmodels}
should be widely separated; 
the inflaton can only move below the horizontal solid line (at
$\phi=R_A\sqrt{n_A T_3}$) which is well below $\phi= \sqrt{2}\mpl$.
This excludes the large field models where $\phi \gtrsim \mpl$, and
opens up a wide region in the middle of the parameter space where
there is no inflation. This is also why 
random-walk eternal inflation in KKLMMT model within the throat
is excluded in \cite{Chen:2006hs},
the tensor mode in brane inflation is unobservable
\cite{Baumann:2006cd},
and the ``intermediate
UV models'' are ruled out in \cite{Bean:2007hc}.\footnote{
Note that in the bulk, the geometric conditions described here alone
are not enough to
restrict the scalar field to be
sub-Planckian. For example, for a toroidal compactification with
$L_i=l_i m_s^{-1}$ $(i=1,\dots,6)$, consider an irregular shape
$l_1>g_s^{-1} l_2 \cdots l_5$. 
Additional consistency requirements are necessary, e.g.
how to maintain the shape of the
potential over $\Delta \phi$ of 
Planckian size while it is expected to vary over the string
scale.}
The earlier statement that the antibrane tension alone is not large 
enough to drive DBI inflation in UV models
can be justified as well by estimating
$H_{\rm anti}\Delta t \approx \sqrt{n_A T_3} R_A h_A/\mpl \ll
1$, where $H_{\rm anti}$ is the contribution from antibrane tension 
and $\Delta
t \approx R_A h_A^{-1}$ is the time scale that the branes spend traveling
down the throat.

Having considered this inflationary phase diagram, we can now restrict
the parameter space by comparing it to observations.
In the UV model, we saw from the above discussions that there
is a clean separation of slow-roll and DBI inflationary phases after
the geometric conditions (\ref{Mplcond})-(\ref{phicond}) are applied. The
slow-roll region is the KKLMMT model \cite{Kachru:2003sx}
and is compatible
with the current observations \cite{Bean:2007hc,Lorenz:2007ze,Battye:2007si}. 
In this model, the inflaton mass may be adjusted to fit the spectral
index. The running of spectral index and 
non-Gaussianities are unobservable if the potential is featureless.
The tensor mode is also unobservable.
The DBI region is the STA model
\cite{Silverstein:2003hf,Alishahiha:2004eh}. 
It predicts large non-Gaussianities with the estimator
\bea
|f_{NL}^{\rm eq}| \approx 1.3 \frac{p^2 \mpl^4}{\phi^4} ~,
\label{UVfNL}
\eea
(where $p = m/(\sqrt{6}\mpl/\sqrt{\lambda_A})\gg 1$,) 
together with a possibly-observable tensor mode $r\approx
5/(p\sqrt{f_{NL}})$ \cite{Alishahiha:2004eh}.
From the constraints (\ref{phicond}) and (\ref{UVcond2}), one can see
that, for one brane $n_A=1$, $|f_{NL}| \gtrsim p^2 (N_A/a_A)^2$ 
which cannot fit the observations
\cite{Bean:2007hc,Chen:2005fe,Baumann:2006cd}.
One way to increase the field range is to increase the number of the
inflaton branes.
Here we emphasize another constraint coming from the relativistic 
probe brane backreactions discussed in
\cite{Silverstein:2003hf,Chen:2004hu}.
In order to treat the mobile inflaton branes as probes of the warped
background, $N_A/a_A \gg n_A \gamma$ is
required. 
Namely, the energy scale of the mobile branes cannot exceed the source
of the warped background.
On the other hand,
combining (\ref{phicond}), (\ref{UVcond2}) and (\ref{UVfNL}), we have
$|f_{NL}| \gtrsim 1.3p^2N_A^2/(a_A^2n_A^2)$. Using the relation
$|f_{NL}|\approx 0.32 \gamma^2$, these two requirements lead to 
$p\ll 0.5$, which is a contradiction. Note that this conclusion 
is independent
of the value of $n_A$ and before any comparison with data is
made.\footnote{Considering wrapped branes
\cite{Kobayashi:2007hm,Becker:2007ui} effectively
interprets $n_A$ in a different way, so it should be subject to the
same conclusion discussed here.}
However we should note that this inconsistency appears when the STA
model is embedded in the warped compactification of the GKP type, so
it remains a viable field-theoretic model, and looking for other
UV embeddings becomes an interesting question.

\subsection{IR models}\label{Sec:IR}
In the IR models, branes are started from the IR side of a throat
(denoted as the B-throat) and roll toward the UV side under the moduli
potential
\bea
V(\phi) = V_0 - \half m^2 \phi^2 ~.
\label{IRpotential}
\eea
The origin of $\phi$ is at the tip of the
throat (\ref{throatmetric}), which can be realized for example if
the tip of the throat is an orbifold fixed point.
The Coulomb attraction from antibranes in other throats 
is neglected here unless $m^2/H^2$ is very small.

\begin{figure}[t]
\begin{center}
\includegraphics[scale=0.35]{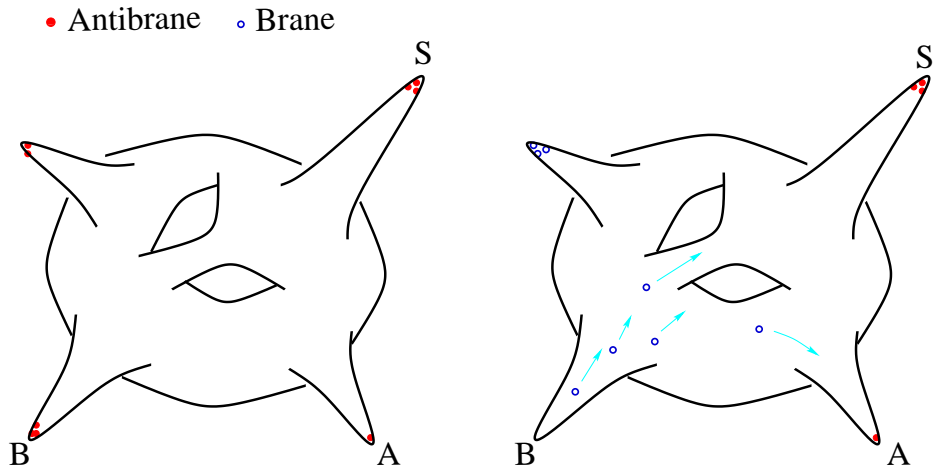}
\end{center}
\caption{\small Multi-throat brane inflation scenario. In the first
figure, antibranes are settled down in throats. In the second figure,
in some throats antibranes annihilate fluxes and generate branes. For
a throat with tachyonic brane moduli, branes fall out and settle down
somewhere else, triggering either IR or UV models of brane inflation.}
\label{Fig:multithroat}
\end{figure}

The IR model can arise in the following scenario
\cite{Chen:2004gc} (illustrated in Fig.~\ref{Fig:multithroat}): 
At the beginning,
antibranes are naturally attracted to and settle down at 
the end of various throats
induced by fluxes. However they are semi-stable at most, and will
eventually annihilate against some fluxes \cite{Kachru:2002gs}. 
The end products are many branes. As mentioned, unlike antibranes, 
branes experience no potential if the extra
dimensions are not compactly stabilized. 
But for realistic inflation models in string
compactification, their moduli space is lifted. If the mass term 
is tachyonic as in (\ref{IRpotential}) for a B-throat, 
these liberated 
branes will roll out.
The inflationary energy $V_0$ is provided by
longer-living antibranes in other throats (denoted as the A-throats) 
or in the bulk. Shorter A-throats give more dominant contributions to
$V_0$. 
These antibranes eventually get annihilated by some of the inflaton
branes.
The annihilation products will naturally heat low mass-scale
sectors in case of tunnelling reheating \cite{Chen:2006ni}, 
such as branes residing in very long throats or in a large
bulk.

\begin{figure}[!ht]
\begin{center}
\includegraphics[scale=0.5]{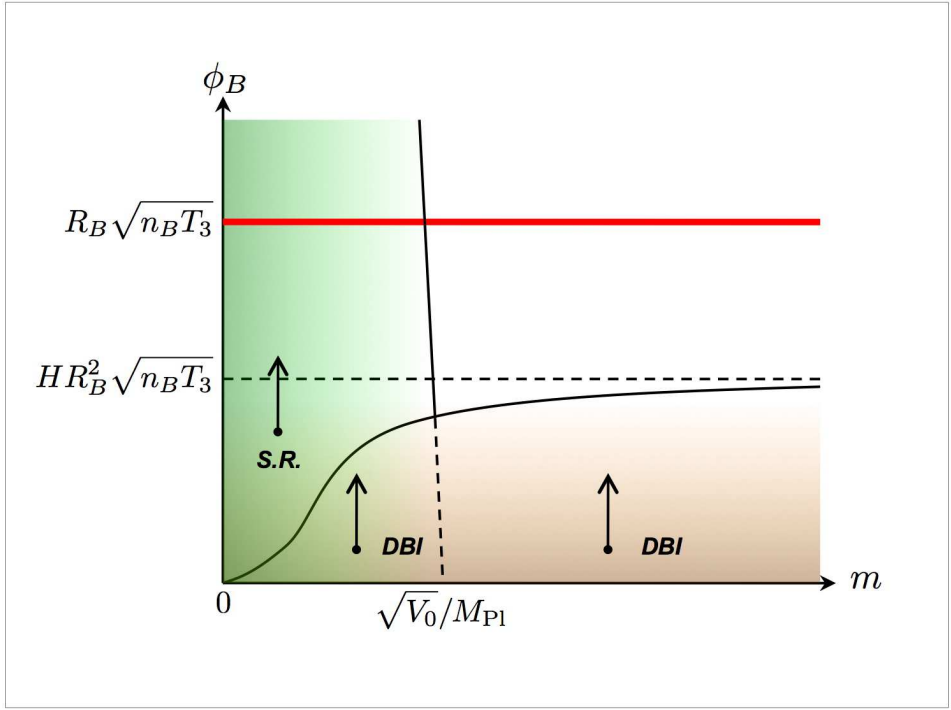}
\end{center}
\medskip
\caption{\small 
The inflation phase diagram for IR models. The notation used
here is the same as in Fig.~\ref{Fig:UVmodels}. The vertical line at
$m=\sqrt{V_0}/\mpl$ corresponds to $|\eta_V| = \beta/3 =1$.
The unshaded region may support a certain amount of non-relativistic
fast-roll inflation.}
\label{Fig:IRmodels}
\end{figure}

In the absence of the warped space, 
the $|\eta_V| \equiv \mpl^2 |V''|/V=1$ line is
\bea
\phi^2 = -2\mpl^2 +\frac{2V_0}{m^2} ~,
\eea
which corresponds to the vertical line (at $m=\sqrt{V_0}/\mpl$) in
Fig.~\ref{Fig:IRmodels}. Slow-roll inflation occurs when $|\eta_V|<1$
which is the region to the left of this line.

In the presence of the warped space, DBI inflation can be triggered
even if the slow-roll condition is not satisfied. 
Ref.~\cite{Chen:2004gc,Chen:2005ad} show that, for $\beta \equiv
m^2/H^2 \sim |\eta_V| \gtrsim 1$, DBI inflation happens if 
\bea
\phi < H R_B^2 \sqrt{n_B T_3} ~.
\eea 
Here, an important difference from the STA model condition
(\ref{UVmcond}) is that, because the inflationary energy $V_0$ is
provided by antibranes in other throats instead of the moduli
potential itself, the shape of the potential that can achieve inflation
becomes rather flexible. Of the special interest is the generic case
$m^2\sim H^2$.
Even for $\beta < 1$ when slow-roll inflation is possible, the
speed-limit provided by the warped space cannot be neglected if the
inflatons are started from the region \cite{Chen:2005ad}
\bea
\phi < \beta H R_B^2 \sqrt{n_B T_3} ~.
\eea
This implies that the DBI and slow-roll phases can be 
smoothly connected
by an intermediate region in this corner of the parameter space.
Overall, the DBI inflation phase stays below the horizontal
curve stretching from the origin to $\phi=H R_B^2 \sqrt{n_B T_3}$ in
Fig.~\ref{Fig:IRmodels}.
The DBI region always stays well below the maximum inflaton extension
(the horizontal solid line at $\phi=R_B \sqrt{n_B T_3}$), since
\bea
\frac{H R_B^2 \sqrt{n_B T_3}}{R_B\sqrt{n_B T_3}} = H R_B 
\lesssim \frac{\sqrt{n_A} h_A^2}{\sqrt{N_B/a_B}} \ll 1 ~,
\label{HRcond}
\eea
where 
\bea
V_0 = 2 n_A h_A^4 T_3 ~,
\label{V0Def}
\eea
and (\ref{Mplcond}) (\ref{Rcond}) are used, 
$n_A$ being the number of antibranes that get
annihilated.
We see that IR DBI inflation is completed
in a very small region at the tip of the B-throat.
Unlike the UV DBI phase, the condition (\ref{phicond}) is
automatically satisfied.
This also justifies the small field expansion in (\ref{IRpotential}).

We have treated the Hubble parameter $H$ as a constant. This can
be verified using the geometric constraints (\ref{Mplcond}) and
(\ref{Rcond}), because in the IR model the potential drop $\Delta
V$ during inflation, estimated very conservatively, satisfies
\bea
\frac{\Delta V}{V_0} \lesssim \frac{m^2 R_B^2 n_B T_3}{2 V_0}
\sim \frac{\beta n_B}{N_B/a_B} ~.
\eea
As long as 
\bea
\beta \ll N_B/(a_B n_B) ~,
\eea
the inflationary energy is approximately a constant.

Therefore we see that in the IR models, inflation can occur for a large
range of the mass parameter,
\bea
0<m^2/H^2 \ll N_B ~,
\label{mrange}
\eea
around the generically expected magnitude $m\sim H$.
The requirement is to start the inflatons from a small enough
$\phi_B$. 
In terms of the flux compactification
this is easy to achieve, because the minimum warp factor is given by
the flux numbers $M$ and $K$ 
in an exponential form \cite{Giddings:2001yu}
\bea
h_{\rm min} \sim \exp(-2\pi K/3M g_s) ~.
\label{hmin}
\eea
We emphasize that non-trivial constraints come from various
back-reactions that
cut off the IR regions of a throat. These include the back-reaction from
the 4-d inflationary background \cite{Chen:2005ad,Chen:2006ni} 
and the back-reaction from
the relativistic inflaton
branes \cite{Silverstein:2003hf,Chen:2004hu}. 
The former is generally (for $\beta \sim 1$) more important
than or comparable to the latter depending
on the number of inflaton branes.
It is estimated \cite{Chen:2005ad,Chen:2006ni} to cut off the throat at
$\phi \sim HR_B^2 \sqrt{n_B T_3}/\sqrt{N_B}$. 
This determines the maximum
number of $e$-folds achievable by the DBI inflationary
phase,\footnote{This constraint is equally important for UV DBI
models.}
$N_{\rm tot}^{\rm DBI} \sim \sqrt{N_B}$.
A more detailed understanding of this backreaction is important.

It is worth pointing out that, in terms of model building, the most
important 
difference between the IR and UV DBI models is not whether branes are
started from the IR or UV side of a warped space. It is the
independence between the inflaton speed-limit and the inflationary
energy, which allows a flexible shape of potential.
We have seen that this naturally happens when
branes are moving out of a throat, with inflationary energy provided by 
antibranes in other throats. Just in terms of field theory, even in
the UV models, such an independence can be achieved by demanding the
constant term $V_0$ in the potential (\ref{UVpot}) to be independent of
the A-throat warp factor, for example by a hybrid of a different field
around $\phi=0$ to suddenly end inflation. 
The question is then how to realize it naturally in
string models.

Having considered the phase diagram, we would like to first
restrict the parameter
space by comparing the predictions of the model with observational
data, and then make predictions for future
observations. This will be the main focus for the 
rest of the paper. We close this subsection with a few comments on the
setup of the model.

Firstly, as we demonstrated in Fig.~\ref{Fig:UVmodels}, for
UV models there is no
inflation around $\eta_V \sim 1$ and a large parameter space beyond
that. So for IR models, it is reasonable to consider the simplest
case where, after 
branes come out of the
B-throat, there is no significant amount of additional 
inflationary $e$-folds if
they roll through the bulk or 
enter another A-throat to annihilate antibranes there. This simplest
possibility represents a fairly generic class of models.

Secondly, more realistic throats such as the Klebanov-Strassler throat
\cite{Klebanov:2000hb}
have a scale-dependent charge. The characteristic scale $R$ decreases
slowly towards the tip of the throat. Especially the geometry around
the tip region will be significantly different from
(\ref{throatmetric}). For UV models such modifications can be
important because the tip of the throat is the region around which 
the last 60 $e$-folds of inflation happens
\cite{Kecskemeti:2006cg,Shiu:2006kj}.
For IR models, the
situation is opposite. The last 60 $e$-folds of inflation happens away
from the bottom of a throat because generally the total $e$-folds is more
than 60. 
Furthermore the relevant field range is very small. 
Therefore, under these conditions (which will be made more precise in
Sec.~\ref{Sec:Attractor}), we
can ignore both the deformation of the throat geometry 
and the running of the throat charge, and approximate the metric as
(\ref{throatmetric}) with the constant $\lambda$ (or
$R$) being the effective value at the relevant $\phi$ (or $r$).

Thirdly, as we have mentioned, the realistic IR case almost always
involves multiple inflaton branes. It is interesting to see whether
the non-Abelian action plays an important role \cite{Thomas:2007sj}.
In our scenario, after the mobile branes are created in the IR end of
the throat, they all have approximately 
the same radial coordinates and roll in the
radial direction. In the angular directions, we either imagine that 
they are randomly distributed, in which case their average separation
is $h_B R_B/n_B^{1/5}$ (the power of $1/5$ is due to the five-sphere)
which is much larger than the local 
red-shifted string length along the extra
dimensions, $h_B m_s^{-1}$; 
or we imagine that they stick
together and roll with a fixed angular coordinate, which is
different from them forming a higher dimensional brane and expanding
around the center. In both cases, we expect
the leading effects of a large number of branes on density
perturbations
to be well represented by the Abelian action.
As in Ref.~\cite{Chen:2005ad,Chen:2005fe}, 
we will use this approximation in this paper.

Lastly, there is a trivial slow-roll region in IR models if we tune
$\beta$ to be near $\CO(0.01)$. We skip this parameter space in this
paper and start from $\beta \gtrsim \CO(0.1)$.

\subsection{Open questions}\label{Sec:OpenQu}

We list two open questions that are relevant to our
parameterization:

$\bullet$ {\em Construction of potentials:} 
Different parameter regions in
the phase diagrams have different requirements on the inflaton mass.
As mentioned, 
the generic magnitude of the inflaton mass is expected to be of order
$H$, but the actual value can be environmental.
For the slow-roll phases in both UV and IR models, such a mass term
has to be tuned to percent level of the generic value;
for the IR DBI model, although the magnitude of the inflaton mass 
can be of the generic order, it has to be tachyonic, or more generally
the potential has to be repulsive for branes. 
For example, the conformal coupling will give a positive mass-squared
$2H^2$ (through the canonical inflaton dependence
in the K\"ahler potential). 
It is possible that such a contribution gets cancelled 
by others from the
superpotential or K\"ahler potential, 
to order $0.01H^2$ for slow-roll models. 
For IR DBI models, it has to be cancelled to a negative value, although
inflation is insensitive to the magnitude. 
In addition it is easy to see
that, for IR DBI inflation to happen,
the shape of the repulsive 
potential can be much more general than the quadratic form 
\cite{ChenUnpublish}.
For the UV DBI model, the requirement seems to be more restrictive.
The typical potential is expected to vary over $\Delta \phi 
\lesssim m_s \ll \mpl$,
so to have a quadratic form over Planckian size $\Delta \phi$ needs
to be functionally fine-tuned. In addition,
using the geometric conditions (\ref{Mplcond}) and
(\ref{Rcond}), the requirement (\ref{UVmcond}) implies 
\bea
m \gg \mpl/\sqrt{\lambda_A} > \frac{N_A^{1/4}}{g_s^{1/4}a_A^{1/2}
n_A^{1/2}} m_s ~.
\eea
In string constructions, the mass parameters typically arise at
most of order of the
string scale $m_s$, with possible suppressions from factors of the
Planck mass $m_s/\mpl$ and warp factors. So 
$a_A \gg \sqrt{N_A/g_s}/n_A$ is necessary, where $a_A$ is defined
before (\ref{RTDef}) and is due to orbifolding.
The construction of all these potentials is an issue under
active investigation
\cite{Berg:2004ek,Baumann:2006th,Burgess:2006cb,Krause:2007jk,Baumann:2007ah}.

$\bullet$ {\em Background D3-charges:}
As we have seen, to get enough $e$-folds in the DBI model, $N_B\sim 10^4$
is enough. But as we will see later, 
there are interesting parameter spaces in
DBI inflation which require very large D3-charges for the A or
B-throat to fit the magnitude of the density perturbations.
This charge can be as large as of order $\CO(10^{14})$
\cite{Alishahiha:2004eh}. 
In IR models, branes are always generated in a large number
after the flux-antibrane annihilation, and this reduces
$N_B$ to $\CO(10^9)$ \cite{Chen:2005ad}.
In a GKP-type
flux compactification, such a charge should be cancelled by the induced
negative D3-charge of the wrapped D7-branes. This negative charge is
given by the Euler number of the corresponding fourfold in F-theory.
The explicit examples give no more than
$\CO(10^5)$ \cite{Klemm:1996ts}. 
So far it is not clear which of the following
possibilities is true: in terms of the density perturbations, 
DBI inflation is extremely fine-tuned 
or even not viable; a modified construction, e.g.
multiple-dimensional
orbifolding (a large $a_A$), can be engineered; 
a more complete understanding of the
flux compactification can give such numbers; 
or subtleties are involved in the reheating.

The approach in this paper to both issues above is phenomenological. 
By parameterizing and comparing them to experimental data,
we can hopefully learn something useful about string
theory from a bottom-up approach.

\section{IR DBI model}\label{Sec:IRDBI}
\setcounter{equation}{0}
In this section, we summarize the main results and
predictions of the IR model, carrying out numerical calculations
whenever is necessary. In the next section,
we compare them to observational data, constrain microscopic
parameters and make predictions.

\subsection{Attractor solutions}\label{Sec:Attractor}

In regions where various back-reactions to
the warped background are negligible
and the Hubble energy stays below the red-shifted string scale, the 
low-energy dynamics of the inflaton branes 
is described by the DBI-CS action
\bea
S= \frac{\mpl^2}{2} \int d^4x \sqrt{-g} R
-\int d^4x \sqrt{-g} \left( \frac{\phi^4}{\lambda_B}
\sqrt{1+\frac{\lambda_B}{\phi^4} g^{\mu\nu} \partial_\mu \phi
\partial_\nu \phi } - \frac{\phi^4}{\lambda_B} + V(\phi) \right) ~,
\label{DBICS}
\eea
where $V(\phi)$ is given by (\ref{IRpotential}).
The branes start from the tip of the throat and end at the UV end
of the throat $\phi_{\rm end} = R_B \sqrt{n_B T_3} 
= \sqrt{\lambda_B}/R_B$. 
After that some of
them quickly find antibranes and annihilate, diminishing the
cosmological constant $V_0$.\footnote{\label{FT:nAnB} 
Among all the branes rolling out
from the B-throat, only those which annihilate antibranes have significant
contributions to the density perturbations. We will denote this number as
$n_B$. More generally, antibranes that get annihilated can reside in
different A-throats. Due to different warp factors, each annihilated
brane pairs can have different contributions to reheating
energy. These subtleties will only
affect the microscopic interpretation of the parameters (such as $n_A
h_A^4$).}

The dynamics of the inflaton can be approximately 
described by two attractor solutions.
The first is that of the IR DBI inflation. 
This is the phase where the effect of
the speed-limit is important. The inflaton is traveling near the
warped speed of light, and the attractor solution is
\bea
\phi = -\frac{\sqrt{\lambda_B}}{t} + \frac{9\sqrt{\lambda_B}}{2\beta^2 H^2
t^3} + \cdots ~,
\label{attractorDBI}
\eea
where $t$ is chosen to run from $-\infty$ for convenience. 
Recall that for
$0<\beta \ll N_B/(a_B n_B)$,  $H$ is approximately a constant.
This phase ends around $\phi \sim H \sqrt{\lambda_B}$ ($t\sim -H^{-1}$)
for $\beta \gtrsim 1$ 
and $\phi \sim \beta H \sqrt{\lambda_B}$ ($t \sim -\beta^{-1} H^{-1}$) 
for $\beta < 1$. The inflationary $e$-folds as the function of $\phi$ 
can be estimated as
\bea
N_e^{\rm DBI} \approx \frac{H\sqrt{\lambda_B}}{\phi} - \beta^{-1} 
~.
\label{NeDBI}
\eea
Here we have incorporated both the case $\beta < 1$ and
$\beta \gtrsim 1$ by adding the term $\beta^{-1}$; 
this correction is negligible for $\beta\gtrsim 1$, and
the validity of (\ref{attractorDBI}) requires $N_e^{\rm DBI} \gg
\beta^{-1}$.

The second attractor solution 
describes the nonrelativistic rolling where the inflaton velocity
stays far below the speed-limit. In this limit the equation of motion
\bea
\ddot \phi + 3H\dot \phi + \partial_\phi V(\phi) =0
\eea
has the following attractor solution
\bea
\phi = \phi_0 e^{\alpha\beta H (t-t_0)/3} ~,
~~~~~ \alpha = \frac{-9+\sqrt{81+36\beta}}{2\beta} ~.
\label{attractorNonrel}
\eea
The consistency condition that the inflaton velocity is
non-relativistic $\dot \phi \ll \phi^2/\sqrt{\lambda_B}$ 
requires that $\phi \gg \alpha \beta H
\sqrt{\lambda_B}/3$.  This phase is smoothly connected to the previous
DBI phase.
The total number of inflationary $e$-folds provided by this period
is given by
\bea
N_{\rm tot}^{\rm NR} \approx
\frac{3}{\alpha \beta} \ln \phi \Bigg|_{\alpha \beta
H\sqrt{\lambda_B}/3}^{\sqrt{\lambda_B}/R_B} 
\approx \frac{3}{\alpha \beta} |\ln H R_B| ~.
\label{NNRtot}
\eea
We emphasize that this nonrelativistic rolling region is slow-roll
only if $\beta \ll 1$, while the above formulae are valid even if this
condition is not satisfied. 
For example in Fig.~\ref{Fig:IRmodels} at around $\beta\sim 1$ ($m\sim
\sqrt{V_0}/\mpl$), after the DBI phase it takes time for branes to go
through the lightly-shaded region till it reaches the end of the
throat. This is because the branes are originally very close to the
top of the potential. It provides a certain amount of $e$-folds typically
not corresponding to the scale of the CMB.
These additional non-relativistic
non-slow-roll inflationary $e$-folds is also interesting to us, because
it affects the relevant $e$-folds in the DBI phase, 
\bea
N_e=N_e^{\rm DBI} +
N_{\rm tot}^{\rm NR} ~, 
\label{NeSum}
\eea
and hence predictions for observations.

\begin{figure}[t]
\begin{center}
\includegraphics[scale=0.8]{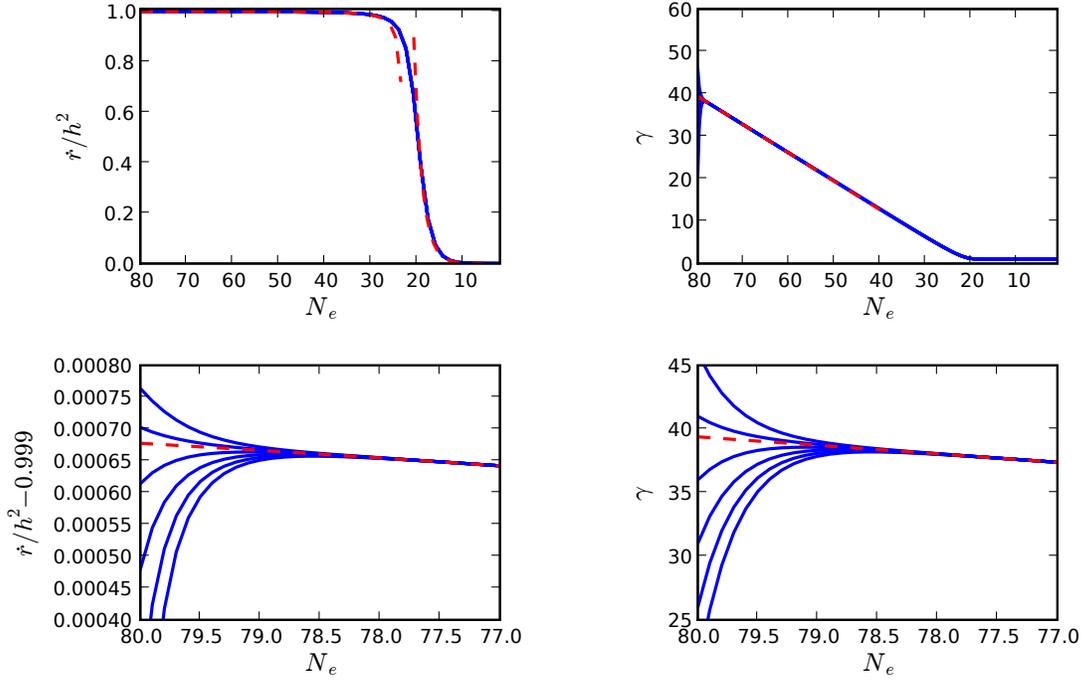}
\end{center}
\caption{\small Attractor solutions and numerical results. 
The dashed lines are the analytical attractor solutions. 
The solid lines are numerical solutions
with different initial velocities.
The upper-left panel shows the evolution of the ratio of the 
inflaton-velocity $\dot r$ to the warped-speed-of-light $h^2$. 
The two dashed lines 
are DBI and non-relativistic rolling, respectively. 
The upper-right panel shows the evolution of the Lorentz factor
$\gamma$.
The lower panels are the blow-ups of the upper panels.
The parameters are $\beta=2$, $N_B=10^{9}$, $n_B=10^{5}$,
$m_s g_s^{-1/4}= 10^{-6} \mpl$, $n_A h_A^4 = 1$. 
In the simulation, branes are started at $h_B=2.9 \times 10^{-7}$.}
\label{Fig:attractor}
\end{figure}

In Fig.~\ref{Fig:attractor} 
we demonstrate numerical results and show that the
two attractor solutions (\ref{attractorDBI}) and
(\ref{attractorNonrel}) give good analytical approximations for the
inflaton dynamics. Any initial angular motions will also be damped out
due to the Hubble friction, because the inflaton potential considered
here has only radial dependence.

\medskip
\noindent {\it Klebanov-Strassler throat}
\medskip

In this paper, 
we use the $AdS_5$ geometry with a length scale $R$ to
represent the warped space. The details of the geometry can be
different for more realistic cases. We expect our example 
to capture the main
properties of the model, and to 
be a good approximation for a certain generic parameter space.
Let us consider, for example, the KS throat,
\bea
h(r)^{-4} &=& \frac{27\pi g_s \alpha'^2}{4r^4}
\left[ N_{\rm tot} + \frac{3g_s M^2}{2\pi} 
\left(\ln\frac{r}{r_{max}} +\frac{1}{4} \right) \right]  
\nonumber \\
&\equiv& \frac{R_l^4}{r^4} ~,
\eea
where we have defined a running $R_l$,
\bea
R_l^4 = \frac{27\pi}{4}g_s \alpha'^2 N_{\rm eff} ~,
\eea
with $N_{\rm eff} \equiv MK_{\rm eff} 
= M(K_{\rm tot} -l)$ for 
\bea
r=r_{max} \exp (-2l\pi/3g_s M) ~.
\label{rl}
\eea
So instead of the parameter $N_{B}$, here we have the parameter $M$.
The effective $N_{\rm eff}$ and $K_{\rm eff}$ are now functions of $r$ or $l$.
From (\ref{rl}) we can estimate that, during IR DBI inflation,
$\Delta l \approx g_s M$. Therefore as long as
\bea
K_{\rm eff} \gg g_s M ~,
\label{Keffcond}
\eea
we can neglect the running of $N_{\rm eff}$. The $N_{B}$ in our analyses
thus represents the $N_{\rm eff}$ in a small region of $r$
relevant for IR DBI inflation. The condition (\ref{Keffcond}) is
most easily satisfied by having a small $g_s$. In addition, since
the WMAP window is only a few $e$-folds, which corresponds to $\Delta l
\approx 3g_sM/N_e^{\rm DBI}$, we only need $K_{\rm eff} > g_sM$ to
approximate $N_{\rm eff}$ as a constant in this window. In this case, 
the running of
$N_{\rm eff}$ only slightly affects the total DBI $e$-folds, and hence the
relation in (\ref{NeSum}). Another difference between the KS throat
and the geometry we use is that, in the latter,
the UV edge of the warped space is cut off and glued to the bulk 
at $R$, while in the
former it is given by an independent parameter $r_{max}$. This does
not cause too much difference in the analyses, since the
non-relativistic fast-roll inflation mostly happens near the top of
the potential; hence $N^{\rm NR}_{\rm tot}$ is insensitive to the
cutoff in generic cases.

\subsection{Power spectrum}\label{Sec:Power}
We first look at the density perturbations in the DBI phase. Its
amplitude is given by the usual formula
\bea
P_k = H^2 \delta t^2 ~,
\eea
where $\delta t$ is the position-dependent time delay caused by the
frozen quantum fluctuations of the inflaton, $\delta \phi = H/2\pi$.
In this phase we approximate the inflaton zero-mode 
velocity as the speed of
light, $\dot \phi= \phi^2/\sqrt{\lambda_B}$. So we have
\bea
P_k = \frac{H^4}{4\pi^2 \dot \phi^2}
\approx \frac{(N_e^{\rm DBI} )^4}{4\pi^2 \lambda_B} ~.
\label{PkDBI}
\eea
If this is responsible for $P_k \approx 23\times 10^{-10}$, we need
$\lambda_B \sim 10^{13}$. Since the number of branes
created after the flux-antibrane annihilation can be as large as
$\CO(\sqrt{N_B})$, this requires $N_B \gtrsim 10^9$.
The formula (\ref{PkDBI}) 
can be derived rigorously using the formalism of
Garriga and Mukhanov \cite{Garriga:1999vw}, 
where we can see that the main difference from
the slow-roll case is the development of the sound speed $c_s$ 
on the world-volume of the
inflaton branes. This shrinks the Hubble horizon by a
factor of $c_s$. The underlying physics can be most easily understood
in the view of an instantaneous co-moving observer with the brane
\cite{Chen:2004gc,Chen:2005ad}. 
For this observer the
Hubble expansion rate is increased by the Lorentz factor
\bea
\gamma = 1/c_s =1/\sqrt{1-\lambda_B \dot \phi^2/\phi^4} 
\approx {\beta N_e^{\rm DBI} \over 3}
\label{gamma}
\eea
due to the relativistic time dilation. 
In the last step of (\ref{gamma}), we have used the IR model solution
(\ref{attractorDBI}) and (\ref{NeDBI}).
This Hubble parameter leads to a
horizon of size $c_s H^{-1}$, which lies orthogonal to
the brane velocity and hence appears the same to the lab observer
(i.e.~the observer that does not move with the branes).

In this model it is very important to realize the validity condition
for the
field theory analyses of density perturbations, and make
estimates for the density perturbations when the field
theory analyses break down
\cite{Chen:2004gc,Chen:2005ad,Chen:2005fe}. 
There are the following several
interesting regions as we extend the inflaton {\em back in time} 
towards the IR side of the warped space.

Firstly, open strings on the inflaton brane will be created when 
the Hubble
energy density $\gamma^4 H^4$ for the moving observer becomes 
larger than the red-shifted 
brane tension\footnote{If we replace the brane
tension with the string scale $m_s^4$, we have an extra factor of
$g_s^{1/8}$ in (\ref{Nc}).} $h_B^4 T_3=\phi^4/n_B\lambda_B$. 
Using (\ref{NeDBI}) and (\ref{gamma}), this happens at the critical
$e$-fold
\bea
N_c \sim
\frac{\lambda_B^{1/8}}{\beta^{1/2} n_B^{1/8}} 
\sim \frac{ N_B^{1/8}}{ \beta^{1/2}} ~.
\label{Nc}
\eea
Another observation that also indicates that we cannot
naively extend the field-theoretic results too far down the IR side of
the throat,
is to look at the region $N_e > N_c$, where
the brane fluctuations in the transverse directions become
superluminal. This is impossible. The reason that such a superluminal
speed even occurs under the DBI action can be understood as
follows. When we calculate the primordial fluctuations, in the first
step, the source of
such fluctuations is the uncertainty principle, which {\it a priori} 
does not necessarily 
respect the speed-limit if we only consider the scalar field. 
In the next step, 
the later evolution of such fluctuations is governed
by the DBI action and always follows the causality constraint. 
The superluminal fluctuation
speeds to which we just referred come from 
the first step, if we naively extend the field
theoretic calculation to the regions beyond (\ref{Nc}).

Secondly, closed strings will be created when the Hubble energy
$H^4$ for the lab observer becomes larger than the red-shifted brane
tension. This happens at
\bea
N_e \sim \lambda_B^{1/4}/n_B^{1/4} ~.\label{closedNe}
\eea

Finally, when the closed string density created by the Hubble expansion
overwhelms the source (fluxes or branes) for the warped geometry,
the warped space gets cut off. As mentioned, this
back-reaction 
determines the maximum number of inflationary $e$-folds in the DBI phase,
\bea
N_{\rm tot}^{\rm DBI} \sim {\lambda_B^{1/2}}/{n_B^{1/2}} ~.
\label{Ntot}
\eea

It is important to note that 
the zero-mode dynamics of the inflaton are still valid as long
as $N_e < N^{\rm DBI}_{\rm tot}$, since it only relies on the
existence of the speed-limit and therefore on the condition (\ref{Ntot})
at which the warped space gets cut off.\footnote{
It will be interesting to understand better how branes move through a
gas of strings and graviton KK modes, whose effects are ignored here.
}
In addition, the strings and graviton KK-modes are only created in the
tip of the throat and have energy density $\CO(H^4)$. It does not
backreact significantly on the Hubble expansion.
However, the field-theoretic calculation of the 
density perturbation is no longer
valid if $N_e \gtrsim N_c$, since not only scalar fields but also open
strings will be created. 

While a rigorous treatment is currently unavailable, 
there are a couple of ways to estimate the density perturbations in
this situation \cite{Chen:2005ad,Chen:2005fe}.
(We shall make the estimates more quantitative in Appendix
\ref{Sec:Width}.)
We can estimate the part of energy that goes
into scalar fluctuations to be saturated when the Hubble temperature
reaches the brane tension $\phi^4/n_B \lambda_B$ at (\ref{Nc}).
Further relative 
increase of the Hubble energy excites strings and branes. The
stringy excitations 
will be diluted by the exponential spatial expansion after the Hubble
energy drops below the brane tension as branes move to the UV side, 
in the same way that the
inflation dilutes relic densities. Only the scalar fluctuations are frozen
and later translated into the position-dependent time-delay for the
reheating. 
For the moving observer, the scalar field energy density is $(\delta
\phi)_{\rm mov}^2 \gamma^2 H^2 \sim \phi^4/n_B\lambda_B$ and for the lab
observer $\delta \phi = \delta \phi_{\rm mov}/\gamma$.
This estimate leads to $\delta \phi =
\sqrt{\lambda_B}H/ \sqrt{n_B} (N_e^{\rm DBI})^2
\gamma^2$.  So for $N_e > N_c$, we estimate
\bea
P_k = H^2 \frac{\delta \phi^2}{\dot \phi^2} 
\sim \frac{324\pi^2}{n_B \beta^4 (N_e^{\rm DBI})^4} ~.
\label{Pklarge}
\eea
This is also the result that we will get by 
looking at the transverse fluctuation speed of each brane. 
The field-theoretic analyses lead to the fluctuation speed (for the moving
observer), $\dot r_{\rm mov} = \delta r/\delta t \approx
{\gamma H \over \sqrt{T_3}}/(\gamma H)^{-1}$, which is the fluctuation
amplitude divided by a Hubble time. This velocity reaches the warped
speed of light precisely around (\ref{Nc}). Above the phase
transition, we assume the fluctuation speed saturates the warped speed
of light $h^2$, so $\delta r_{\rm mov} \approx h^2 (\gamma H)^{-1}$.
The position-dependent time-delay is then 
\bea
\delta t_{\rm lab} \approx 
\frac{\delta r_{\rm mov}/\dot r_0}{\gamma \sqrt{n_B}} \approx
\gamma^{-2} H^{-1} /\sqrt{n_B} ~,
\label{deltatlab}
\eea
where $\dot r_0 = h_B^2$ is the zero-mode brane speed, and the factor
$1/\sqrt{n_B}$ is due to the
reduction of the root-mean-square of the fluctuations by the
superposition of $n_B$ independent branes.
Eq.~(\ref{deltatlab}) reproduces (\ref{Pklarge}).

\subsection{Regional large running of spectral index and phase
transition} \label{Sec:Index}
From the last subsection, the spectral index is 
\bea
n_s-1 = \frac{d \ln P_k}{d \ln k} \approx -\frac{4}{N_e^{\rm DBI}} 
\label{nssmall}
\eea
for $N_e^{\rm DBI} < N_c$, and
\bea
n_s-1 \sim \frac{4}{N_e^{\rm DBI}} 
\label{nslarge}
\eea
for $N_e^{\rm DBI} > N_c$.

The most interesting information from 
(\ref{Nc}) is its smallness due to
the power $1/8$. For $\lambda_B \sim
10^{13}$, $\lambda_B^{1/8} \sim \CO(100)$, which already makes $N_c$
interestingly small. 
Considering the more realistic multi-brane case $n_B \lesssim
\sqrt{N_B}$ leads to even smaller $N_c$ of order $\CO(10)$.
So such an interesting phase transition
may well have occurred within our CMB scale.
Another interesting property is the fact that Eq.~(\ref{Nc}) is
independent of the inflationary energy scale and the local warp factor
of the inflaton branes, so it will be a rather generic prediction of
the IR DBI models.\footnote{\label{Ft:UVphasetran} In the STA model
\cite{Silverstein:2003hf,Alishahiha:2004eh}, 
when the branes move towards the IR side of the warped space,
the Hubble energy drops
linearly as $\phi$, $H \approx m\phi/\mpl$, the same as the warp factor. 
Comparing the relativistic Hubble energy $\gamma H \approx \gamma
m\phi/\mpl$ with the red-shifted brane tension $T_3^{1/4} h_A =
\phi/(n_A \lambda_A)^{1/4}$,
the phase transition happens for
$\gamma > \mpl/(m n_A^{1/2} N_A^{1/4})$,  
in which the spectral index transitions from
$n_s-1\approx 0$ at large scales to $n_s-1 \sim -8/p$ at small
scales. (In this footnote, 
we are ignoring the issue of UV embedding discussed in
Sec.~\ref{Sec:UV}.)
}

It is very important how sharp this
transition is in terms of $e$-folds. In Appendix \ref{Sec:Width}, 
we give an
estimate of the transition width based on the following approach.
For the familiar case of field-theoretic density perturbations, 
the super-horizon perturbations can
be understood as being generated by the random walk of the transverse brane
fluctuations within a Hubble time before the modes exit the
horizon. Such a random walk velocity is given by the Hubble energy,
and is non-relativistic. As we have discussed, during
or above the stringy phase transition, the main
difference is that the Hubble energy is comparable to or 
exceeds the rest mass of the
brane in a Hubble-size patch. As a consequence the brane
fluctuation speed becomes relativistic.
We therefore use the same
physical picture underlying the familiar theory, 
but generalize it relativistically to estimate the behavior of the
density perturbations across the phase transition. 
The result is given in (\ref{PkallA}).

It is worth noting that this scenario has marked differences to
several other cases commonly discussed in the literature. 
Firstly, this model has a scale-varying running of $n_s$, in
contrast to the commonly investigated empirical ansatz, 
where the running of $n_s$ is assumed to be constant.
Here the large running of $n_s$ is only
regional, principally when $N_e^{\rm DBI}\lesssim N_c$.
Secondly, this scenario can generate  large running, in contrast to
most slow-roll scenarios. A standard slow-roll
potential predicts very small running of $n_s$.
A transient, large $dn_s/d\ln k$ 
can be caused by some ``mild features'' in the potential. For example,
for small field slow-roll inflation, to
generate a large transition for $n_s$ from blue to red, 
the mild feature should be a potential shape changing 
from concave to convex.
We study this case in Appendix \ref{Sec:Mild}.
Since the spectral index is still close to one, the slow-roll
parameters for this case should still be at least of order $0.1$. So
such a case predicts unobservable non-Gaussianities. 
Lastly, non-standard choice of vacua may also cause
observable running of spectral index. As we will discuss in Appendix
\ref{Sec:nonBD}, such a running will be oscillatory in the WMAP window
and phenomenologically distinguishable from the phase transition 
in IR DBI inflation.

\subsection{Large non-Gaussianity and small tensor mode}
\label{Sec:nonG}
The three-point function of the scalar perturbation
for general single field inflation models,
where the Lagrangian is an
arbitrary function of $X=-\half g^{\mu\nu} \partial_\mu \phi
\partial_\nu \phi$ and $\phi$, is derived in
Ref.~\cite{Chen:2006nt}.
In the absence of any sharp features \cite{Chen:2006xj}, 
large non-Gaussianities can
arise if the sound speed $c_s \ll 1$ or another quantity
$\lambda/\Sigma \gg 1$ (related
to the third derivative of the Lagrangian with respect to $X$).
This non-Gaussianity is a function of 
three momenta, which are conveniently referred
to as the shape of the non-Gaussianity \cite{Babich:2004gb} 
and the running of the
non-Gaussianity \cite{Chen:2005fe}. 
The former describes its dependence on the shape of
the momenta triangle, and the latter the overall size of the triangle.
In the absence of sharp features, the running is relatively weak, and
the shape has two categories: (1) the ``local shape'' in which the
non-Gaussianity blows up in the squeezed-limit (where one of the
momenta
goes to zero) and takes a minimum value in the equilateral-limit (where
all three momenta are equal); (2) the ``equilateral shape'' in which the
non-Gaussianity vanishes in the squeezed-limit and reaches maximum value
in the equilateral limit.
The primordial non-Gaussianity is considered to be possibly observable
if its estimator $f_{NL}$
is $|f^{\rm loc}_{NL}| >2$ 
or $|f^{\rm eq}_{NL}|>10$
\cite{Hikage:2006fe,Sefusatti:2007ih}, where the superscript ``loc''
refers to the local shape and ``eq'' the equilateral shape. 

For DBI inflation, the result becomes\footnote{The papers
\cite{Maldacena:2002vr,Chen:2006nt} chose an opposite sign convention
of $f_{NL}$ to the WMAP convention
\cite{Komatsu:2001rj,Spergel:2006hy,Creminelli:2006rz}. In this paper,
we quote $f_{NL}$ in the WMAP convention. We thank
Marilena LoVerde and Sarah Shandera for the clarification.}
\cite{Alishahiha:2004eh}
\bea
f_{NL}^{\rm eq} \approx -0.32 c_s^{-2} ~.
\eea
For IR DBI inflation, using the relation (\ref{gamma}), we have
\cite{Chen:2005fe}
\bea
f_{NL}^{\rm eq} \approx -0.036 \beta^2 (N_e^{\rm DBI})^2 ~.
\label{IRfNL}
\eea
The current observational bound is $-256 < f_{NL}^{\rm eq} < 332$
\cite{Creminelli:2006rz}.
Comparing (\ref{UVfNL}) and (\ref{IRfNL}), we see that the
requirements
of the non-Gaussianity bound on the fundamental parameters are quite
different. Furthermore 
the running of non-Gaussianities
for these two cases 
are opposite, as dictated by the background geometry scanned
through by the rolling inflatons.

However, we emphasize that the above results are derived in the regime
where the primordial fluctuations are field-theoretic. Therefore
the results can be different when the stringy phase transition happens.
As we will see, data analysis
suggests that the critical
scale $k_c$ for that transition lies somewhere near the largest
scales. For those smaller scales, it
seems reasonable to assume that the magnitude of Eq.~(\ref{IRfNL}) 
should be smoothly
modified by the stringy corrections.
So in this paper, we will
use this field-theoretic approximation (\ref{IRfNL}) and 
the bound $|f_{NL}^{\rm eq}|<256$.

For DBI inflation, the scalar and tensor perturbations can be
written as follows,
\bea
P_k = \frac{f V^2}{36\pi^2 \mpl^4} ~, ~~~~~~
P_h = \frac{2V}{3\pi^2 \mpl^4} ~,
\eea
where $f(\phi)$ is the background geometry and for our case $f =
\lambda_B/\phi^4$. 
So the tensor to scalar ratio is
\bea
r_{TS} \equiv \frac{P_h}{P_k} = \frac{24 \phi^4}{\lambda_B V} ~.
\label{tensor}
\eea
Because $V$ is almost a constant in this model, the fact that the
scalar and tensor modes have different horizon sizes during
inflation makes no difference to (\ref{tensor}).
From Sec.~\ref{Sec:Attractor},
we see that, at the scale of $N_e^{\rm DBI}$, $\phi \approx H R_B^2
\sqrt{n_B T_3}/N_e^{\rm DBI}$.
So we get
\bea
r_{TS} &\approx& 
\frac{8}{{N_e^{\rm DBI}}^4} (H R_B)^2 \frac{n_B R_B^2 T_3}{\mpl^2} 
\nonumber \\ 
&\lesssim& \frac{1}{{N_e^{\rm DBI}}^4} \frac{n_A h_A^4}{N_B/a_B} 
\frac{n_B}{N_B/a_B} ~.
\label{rbound}
\eea
This is of course consistent with the Lyth bound
\cite{Lyth:1996im,Baumann:2006cd,Lidsey:2007gq}, since the r.h.s.~of
(\ref{rbound}) is just the square of $\Delta \phi/\Delta N_e$ in
Planck units divided by $(N_e^{\rm DBI})^2$, 
as we can see from (\ref{UVcond2}), (\ref{HRcond}) and (\ref{NeDBI}).
To ignore the probe brane back-reactions we need 
$\gamma n_B \ll N_B/a_B$.
So 
\bea
r_{TS} \ll 1/({N_e^{\rm DBI}}^4 \gamma^2) < 10^{-6} ~,
\eea
which is unobservably small. Therefore 
in our data analyses, we always set $r_{TS}=0$.

\subsection{Constraining microscopic parameters}
\label{Sec:Parameters}
In this subsection we identify the set of 
microscopic parameters of the model, and list self-consistency
constraints and the observables.
As discussed in Appendix \ref{Sec:Width}, we can estimate the power
spectrum at all scales across the transition region 
by the following formula,
\bea
P_k = \frac{4\pi^2 v^2 T_3}{\gamma^4 \dot \phi^2} ~,
\label{Pkall}
\eea
where
\bea
v^2 T_3 = h_B^4 T_3 \left[1- \left(1+
\high{ \frac{\gamma^4 H^4}{32\pi^4 h_B^4 T_3} }\right)^{-2} \right]
= \high{ \frac{\phi^4}{n_B \lambda_B}
\left[ 1- \left(1+
\high{ \frac{n_B\lambda_B \gamma^4
H^4}{32\pi^4\phi^4} }\right)^{-2} \right] }
~.
\label{v2T3}
\eea
These equations relate the fundamental parameters to the
observations. 
We will always choose initial position and velocity of
branes so that all the observable scales are within the attractor
solution, namely the total $e$-folds is larger than the minimum
requirement. So these initial conditions will not enter the
observables.
The parameters $V_0$ and $\beta\equiv m^2/H_0^2$ 
determine the scale and shape of the relevant part
of the potential, $\lambda_B/n_B$ characterizes the background
geometry, and $R_B$ tells us where to end the inflation ({\it{i.e.}}~at
$\phi_{\rm end} = \sqrt{32\pi^2/27} \sqrt{\lambda_B}/R_B$).
So these four parameters determine the zero-mode evolution of
the spacetime background and the inflaton dynamics in terms of the
number of inflationary $e$-folds to the end of the inflation.
In particular this determines the evolution of $\gamma$,
$\phi/\sqrt{n_B}$
and $H$ in (\ref{Pkall}) and (\ref{v2T3}). Note that we can write the
factor $h_B^4 T_3$ in (\ref{v2T3}) in terms of $\phi/\sqrt{n_B}$ and
$\lambda_B/n_B= 2\pi^2N_B$, $h_B^4 T_3=(\phi/\sqrt{n_B})^4/(\lambda_B/n_B)$,
so this is also determined. Because the factor $\dot
\phi^2$ appears in the denominator of (\ref{Pkall}), the parameter $n_B$
affects the overall scale of $P_k$, but does not affect the spectral
index once $\lambda_B/n_B$ is fixed.
In conclusion, we have five parameters $\{ \lambda_B,
n_B, R_B, V_0, \beta \}$. Using (\ref{RTDef}) and (\ref{V0Def}), these
parameters are equivalent to five even more fundamental microscopic parameters 
$\{ N_B, n_B, m_s g_s^{-1/4}, n_A
h_A^4, \beta \}$.

We have the following observables:
\begin{enumerate}
\item The amplitude of the power spectrum $P_k
\approx 23 \times 10^{-10}$. 
Through (\ref{PkDBI}) and (\ref{Pklarge})
this roughly determines the order of magnitude of the parameters
$\lambda_B$ or $n_B \beta^4$ depending on whether the pivot point
$N_e$ is smaller or larger than $N_c$.

\item The scale-dependence of $P_k$. This determines the
spectral index and its running. Since the spectral index of this model
has a regional large running, the data will constrain at which scale
($k_c$) such a running happens and which DBI $e$-fold ($N_c$) it
corresponds to. These are then transferred into some delicate
relations with the microscopic parameters, e.g.~(\ref{Ncdef}) and 
(\ref{NeSum}).

\item Non-Gaussianity constraint $\gamma<28$, will mainly
constrain $\beta$ (with some weak dependence on $N_e^{\rm DBI}$).

\item We have the following several consistency relations. First,
a scale $k$ is related to the corresponding
$N_e$ by
\bea
N_e = 65 - \ln \frac{k}{0.002\ {\rm Mpc}^{-1}} + \ln \frac{H_0/\hat c_s}{T_{\rm
reheat}} ~,
\label{Nek}
\eea
where the reheating is assumed to be efficient\footnote{For
single-throat reheating, the brane-antibrane pairs immediately (in terms of
the Hubble time) annihilate and decay into relativistic particles and
start the usual radiation-domination epoch. For tunneling reheating
such as double-throat reheating, (\ref{Nek}) may receive some small
modifications due to a long intermediate matter-domination
epoch \cite{Chen:2006ni}.}
so that $T_{\rm reheat}
= V_0^{1/4}$, and $\hat c_s$ is the sound speed when the mode
$k=0.002\ {\rm Mpc}^{-1}$ crossed the sound horizon. 


Second, according to the multi-throat brane inflationary scenario, 
the maximum number of the inflaton branes is bounded by the flux
number $M$. Since $N_{B} = a_BKM$ and we want to keep the minimum warp
factor (\ref{hmin}) small, we require
\bea
n_B \lesssim \sqrt{N_B/(a_B g_s)} ~.
\label{nBbound}
\eea

Third, 
the geometric constraints (\ref{Mplcond}) and (\ref{Rcond}) give
an upper bound on $m_s g_s^{-1/4}$,
\bea
\frac{m_s}{g_s^{1/4}} 
\lesssim 2^{3/2}\pi^{11/4} a_B^{1/2} \frac{\mpl}{N_B^{3/4}} ~,
\label{msbound}
\eea
where the approximate
numerical factors come from the toroidal compactification, 
$\mpl^2 = \frac{2}{(2\pi)^7} g_s^{-2} m_s^8 V_6$,
which may change for more realistic setups.

Fourth, the warp factor $h_A \le 1$ and $n_A=n_B$, (this is not a
coincidence, see footnote \ref{FT:nAnB}), so
\bea
n_A h_A^4 \le n_B ~.
\eea

Lastly, the inflationary scale and the string scale are both
bounded below by TeV,
\bea
n_A h_A^4 T_3 \ge {\rm TeV}^4 ~,\\
m_s g_s^{-1/4} \ge {\rm TeV} ~.
\eea
(These two bounds turn out to be very weak. Much stronger ones will
arise from the data analyses.)

Note that, there are two other independent 
parameters $g_s$ and $a_B$ that only
appear in the bounds (\ref{nBbound}) and (\ref{msbound}). For
simplicity we do not promote them into free parameters in data
analyses.
We set $g_s=0.1$ and $a_B = 1$ in these bounds.
Reducing $g_s$ and/or increasing $a_B$ may loosen these
bounds and allow some microscopic parameters to take wider ranges.
But this should not change the model predictions qualitatively.

\end{enumerate}

\section{Markov Chain Monte Carlo data analysis}
\label{Sec:MCMC}
\subsection {Methodology} \label{methods}

In order to obtain multi-dimensional parameter constraints 
from cosmological data,  a Markov Chain Monte Carlo (MCMC) approach 
\cite{Christensen:2000ji,Christensen:2001gj,Knox:2001fz,CAMB,LewBri02,Kosowsky:2002zt,Verde:2003ey}
is employed to sample the likelihood surface efficiently. The MCMC  is
used to simulate observations  from the posterior distribution
$P(\alpha|x)$, for a set of parameters $\{\alpha\}$ given an event
$\{x\}$ (which, for us, is the total set of observational data), using
Bayes' Theorem 
\bea \label{eq:bayes}
P(\alpha|x) = \frac{P(x|\alpha) P(\alpha)}{\int P(x|\alpha)
P(\alpha)d\alpha}
\eea
where $P(x|\alpha)$ is the likelihood of the event $x$ given the model
parameters $\alpha$, and $P(\alpha)$ is the prior probability
distribution of obtaining a model parameter value $\alpha$. The MCMC
generates random draws from the posterior distribution that are a
``fair'' sample of the likelihood surface, and from this sample, we
can estimate all the quantities of interest about the posterior
distribution (mean, variance, confidence levels).  

In most cosmological analyses, flat priors, $P(\alpha)$=constant, are
assumed on a set of empirical parameters such as the spectral index
and its running, $n_s$, $d n_s /d \ln k$, and the normalization $A_s$,
of the primordial scalar power spectrum, or its logarithm $\ln A_s$
(for example \cite{Peiris:2003ff,Parkinson:2004yx,
Spergel:2006hy,Bean:2006qz,Bean:2007hc}). It is by no means true,
however, that such constant priors should naturally arise in a
fundamental theory. The effect of priors on constraints on slow-roll
inflation was recently discussed in \cite{Peiris:2006ug}; here we
discuss their role in IR DBI inflationary scenarios. 

Unlike parameters used in an empirical ansatz, the relationships
between the fundamental microscopic parameters and observables are
highly nonlinear and far from transparent. This can make it
problematic for the MCMC to efficiently explore the likelihood
surface, potentially leading to the presence of non-Gaussian posterior distributions: for example, multiple, disconnected
maxima in the likelihood surface, or long, curved degeneracy
directions. In these cases, a proposal distribution for the
microscopic parameters that samples the posterior distribution
efficiently can be very difficult to obtain. In such situations,
instead of directly adopting the microscopic parameters as the
parameters sampled by the MCMC, we find it useful to reparameterize
variables according to the properties of the models. The specific
details will of course be model-dependent, but there are certain
general strategies that one can follow, which we will summarize now. 

\begin{itemize} 
\item Although the full relationships between the observables and
microscopic parameters $\{\alpha\}$ are usually complicated, and in
realistic cases often have to be computed numerically, an isolated
analytical expression for the observationally accessible window
(scales $10^{-4}\ {\rm Mpc}^{-1} \lesssim k\lesssim 1\ {\rm
Mpc}^{-1}$) can be much easier to obtain and be expressed in terms of
an equal or smaller number of effective parameters $\{\theta\}$. 

\item Run a trial MCMC with the effective parameters $\{\theta\}$ with
constant priors, in order to ensure that these parameters have a
relatively simple likelihood surface. This will generally be the case
if the $\{\theta\}$  are chosen to such that the observables of the
model vary roughly linearly with the effective parameters.  

\item The effective parameters $\{\theta\}$ can often provide the
necessary physical intuition to find a reparameterization of the
original microphysical parameters $\tilde\alpha_i (\alpha_i)$, which
have simple (e.g.~linear) relationships to the $\{\theta\}$,
thus have simple enough relationships to the observables such that the
likelihood surface can be effectively explored by standard, robust
MCMC techniques. Ideally, the reparameterization $\tilde\alpha_i
(\alpha_i)$ should be a bijective function in the observable region of
interest. Because the trial MCMC helps ensure the simplicity of the
likelihood surface in the space of $\{\theta\}$, and the new
parameterization ensures that the $\{\tilde\alpha\}$ essentially
travel along the directions of $\{\theta\}$, the likelihood surface in
space of $\{\tilde\alpha\}$ will also be plausibly simple. When
running the full MCMC in the $\{\tilde\alpha\}$ space, any analytical
approximations used in the trial MCMC to compute the observables can
be dropped, and the observables calculated numerically, in order to
prevent modeling uncertainties coming from such approximations
significantly affecting the final constraints. 

\item After obtaining the likelihood surface of the new parameters
$\{\tilde\alpha\}$, transform the likelihood surface of the
$\{\tilde\alpha\}$ to the space of the original parameters
$\{\alpha\}$; the MCMC can also be re-weighted to impose any desired
priors on the $\{\alpha\}$. It must be noted at this stage that the
theory does not predict the prior distribution of the $\{\alpha\}$ and
therefore any prior adopted on this parameter set can potentially be
highly informative. If the data impose a tight-constraint on a given
parameter ({\it{i.e.}} the likelihood is significantly peaked within
the prior) such that the posterior distribution is not very sensitive
to simple forms of adopted prior (such as constant, logarithmic etc),
we will not concern ourselves overly with this point. If a given
``constraint'' is coming primarily from the prior, we will point it
out. 

\item An alternative approach, which is to use complicated sampling
techniques to explore the complex likelihood surface of the original
microphysical parameters $\{\alpha\}$, can often be more
time-consuming as it has to be tuned for each particular problem.

\end{itemize}

Now we will apply this procedure to the IR DBI model.

\subsection {MCMC using microscopic parameters of the IR DBI model} 
\label{Sec:reparam}

First, as detailed in Appendix \ref{Sec:Width}, we notice that the
primordial power spectrum predicted by the IR DBI model can be
approximated by the following analytical form:
\bea
P_k = H^2 \delta t^2 = \frac{324\pi^2}{n_B \beta^4 {N_e^{\rm DBI}}^4}
\left[ 1- \frac{N_c^{16}}{(N_c^8 + {N_e^{\rm DBI}}^8)^2} \right]~.
\label{PkDBIall1}
\eea
This is parameterized by three effective 
parameters: $N_c$, $\ln k_c$, and $n_B
\beta^4$, where $k_c$ is the critical scale near which the stringy
phase transition happens,
\bea
N_e^{\rm DBI} = \ln (k_c/k) + N_c ~.
\label{kcdef}
\eea

After verifying that these three parameters  
appear to have a simple likelihood surface in a trial MCMC
analysis, we relate the five microscopic parameters to these three
parameters through approximate analytical expressions.

The relation to $N_c$ is simple and given by (\ref{Ncdef}),
\bea
N_c = \sqrt{6} \pi^{1/4} \frac{N_{B}^{1/8}}{\beta^{1/2}} ~.
\eea

The $k_c$ is defined as the value of $k$ at $N_e^{\rm DBI} = N_c$.
Using the relations (\ref{NeSum}) and (\ref{Nek}),
we get
\bea
N_c + N_{\rm tot}^{\rm NR} = 65 - \ln \left(\frac{k_c}{0.002\ {\rm Mpc}^{-1}}\right)
+ \ln \left( \frac{H_0}{\hat c_s T_{\rm reheat}}\right) ~.
\eea
Using the approximation (\ref{NNRtot}), expressing $H_0$, $R_B$, $V_0$ in
terms of $N_{B}$, $n_Ah_A^4$ and $g_s/m_s^4$ using Eq.~(\ref{RTDef})
and (\ref{V0Def}), we obtain
\bea
\ln \left(\frac{k_c}{0.002\ {\rm Mpc}^{-1}}\right)
&\approx& 65 - N_c + \ln \frac{1}{\sqrt{6}\pi^{3/4} \hat c_s}
+ \frac{3}{\alpha\beta} \ln \frac{a_B^{1/4}}{\sqrt{6}\pi^{5/4}}
+ \frac{3}{4\alpha\beta} \ln N_{B}
\nonumber \\
&+& \left( \frac{1}{4} + \frac{3}{2\alpha\beta} \right) \ln n_Ah_A^4
- \left( \frac{1}{4} + \frac{3}{4\alpha\beta} \right) \ln
\frac{g_s}{m_s^4} ~,
\label{kcrelation}
\eea
where 
\bea
\alpha\beta = (-9+\sqrt{81+36\beta})/2 ~.
\eea
Here, $\hat c_s$ is the sound speed when the mode $k=0.002$ Mpc$^{-1}$
crosses the sound horizon; we can also approximately 
express it in terms of the five
microscopic parameters. But the detailed expression will complicate
the relation. For our purpose, since $\hat c_s$ varies slowly from $0.01$ to
$0.1$, treating it as a constant should not cause problems for the
reparameterization.

These expressions suggest that the following set may prove to be a
successful reparameterization of the microphysical parameters that can
be effectively explored by MCMC:
\bea \label{eq:reparam}
\tilde\alpha_1 &=& N_c = 
\sqrt{6} \pi^{1/4} a_B^{1/8} \frac{N_{B}^{1/8}}{\beta^{1/2}} ~,
\nonumber \\
\tilde\alpha_2 &=& \frac{3}{4\alpha\beta} \ln N_{B}
~,
\nonumber \\
\tilde\alpha_3 &=& \left( \frac{1}{4} + \frac{3}{2\alpha\beta} \right) \ln
n_Ah_A^4  ~,
\nonumber \\
\tilde\alpha_4 &=& \left( \frac{1}{4} + \frac{3}{4\alpha\beta} \right) \ln
\frac{g_s}{m_s^4}
-\left( 3 \ln \frac{a_B^{1/4}}{\sqrt{6}\pi^{5/4}} \right)
\frac{1}{\alpha\beta}
~,
\nonumber \\
\tilde\alpha_5 &=& \ln n_B \beta^4 ~.
\eea
The relation between these new parameters, $\{\tilde \alpha\}$,
and the effective parameters,
$\{\theta\} = \{N_c, \ln k_c, n_B\beta^4\}$, 
is very clear. Two of them are
identical, and the rest, $\tilde \alpha_2$, $\tilde \alpha_3$, and
$\tilde \alpha_4$, all have approximately linear relationships to $\ln k_c$
through (\ref{kcrelation}).

We adopt the reparametrized microscopic parameters
$\{\tilde{\alpha}\}$ and the standard set of cosmological parameters
$\{\omega_b \equiv \Omega_b h^2$, $\omega_m \equiv \Omega_m h^2$,
$\theta_A, \tau\}$ as the model parameter set sampled by the
MCMC. Here, $\theta_A$ is the angular size of the acoustic horizon and
functions as a proxy for the Hubble constant $H_0\equiv 100h$ km/s/Mpc
or $\Omega_{m}$, and $\tau$ is the optical depth to reionization. The universe is assumed to be spatially flat.
Constant priors are assumed over the previously specified parameter
set $\{\tilde \alpha \}$, 
subject to the microphysical cuts described below. 

For each set of $\{\tilde{\alpha}\}$ sampled by the MCMC, the
relations (\ref{eq:reparam}) are numerically inverted to obtain the
set of microscopic parameters $\{N_B$, $n_B$, $n_Ah_A^4$,
$g_s^{-1/4}m_s$, $\beta\}$. This inversion is bijective in the
parameter ranges of interest. The microscopic parameters are then fed
into a numerical code which is described in detail in Appendix
\ref{Sec:Numerical}. After checking the input parameters for a set of
microphysical conditions which enforce model-building
self-consistency, as described in the Appendix \ref{Sec:Numerical}, 
the code computes the
primordial power spectrum of the curvature perturbation for the model
specified by the input parameters. Input parameters which fail to
satisfy the microphysical cuts are rejected through being assigned
zero likelihood in the MCMC. The primordial power spectrum from this
code is fed to the Boltzmann code CAMB \cite{CAMB}, without
significantly increasing the computational time, in order to calculate
the cosmological observables.  

We use a modified version of the CosmoMC code \cite{LewBri02} to
determine constraints placed on this parameter space by the WMAP
three-year cosmic microwave background data
\cite{Spergel:2006hy,Page:2006hz,Hinshaw:2006ia,Jarosik:2006ib} and
the SDSS Luminous Red Galaxy (LRG) galaxy power spectrum data
\cite{Tegmark:2006az}. We marginalize analytically over the linear
bias factor $b$ and the non-linearity parameter $Q_{\rm nl}$ of the
SDSS LRG data as is done normally in the CosmoMC code. A properly
derived and implemented MCMC draws from the joint posterior density
defined in (\ref{eq:bayes}) once it has converged to the stationary
distribution. We use eight Markov chains and a conservative
Gelman-Rubin convergence criterion \cite{gelman/rubin:1992} on the
eigenvalues of the parameter covariance matrix  to determine when the
chains have converged to the stationary distribution. Then we re-weight the MCMC to switch to constant priors on the
microscopic parameters $\{\alpha\} =\{\log_{10} N_B, \log_{10} n_B, \beta, \log_{10} n_A h_A^4, \log_{10} m_s/g_s^{1/4} \}$.

\begin{figure}[ht!]
\begin{center}
\includegraphics[scale=1]{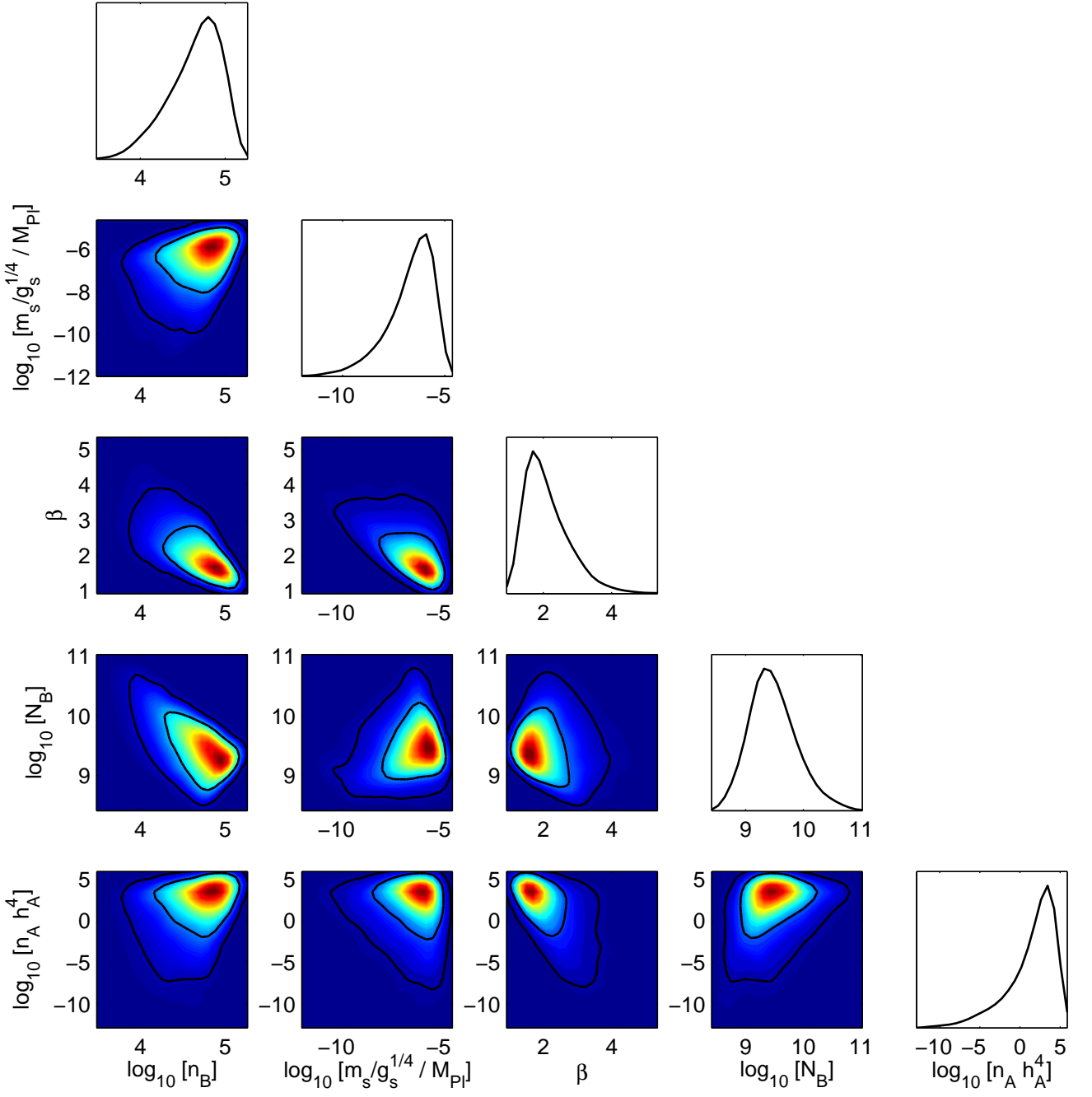}
\caption{\label{Fig:fund_tri} \small
Solid lines show the marginalized 2D-joint 68\% and 95\% probability
contours (off-diagonal panels) and 1D marginalized probability distribution (diagonal panels) for the microphysical IR DBI parameters. The color coding in the off-diagonal panels shows the
marginalized probability density in these 2D parameter spaces, ranging from red
for the highest density to blue for the lowest.}
\end{center}
\end{figure}

\begin{figure}[ht!]
\begin{center}
\includegraphics[scale=0.85]{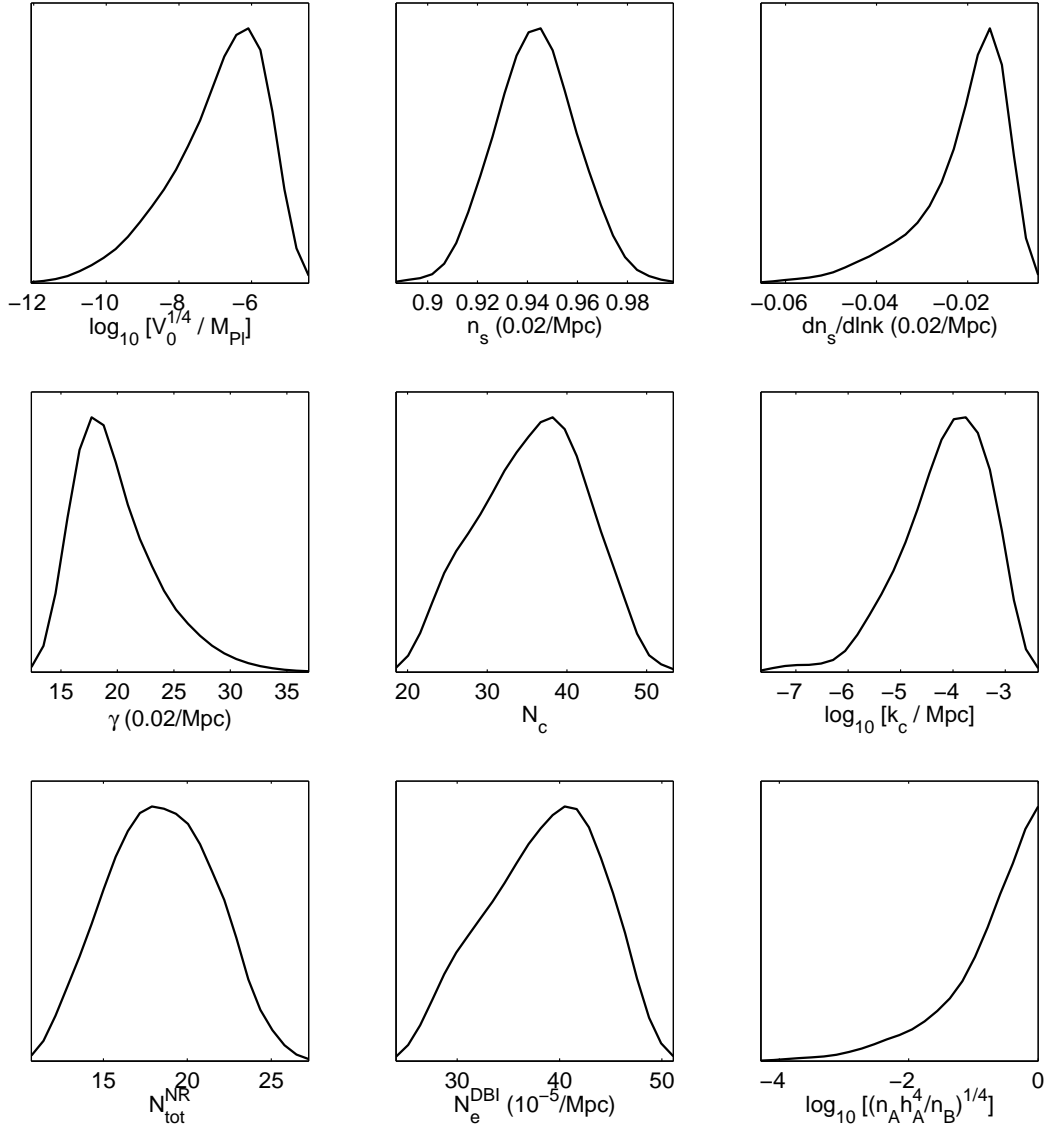}
\caption{\label{Fig:1d}
\small Marginalized posterior probability distribution functions obtained from the MCMC analysis for observables and derived quantities of interest. The functions are normalized such that the area under the curve is one.}
\end{center}
\end{figure}

\begin{figure}[ht!]
\begin{center}
\includegraphics[scale=0.9]{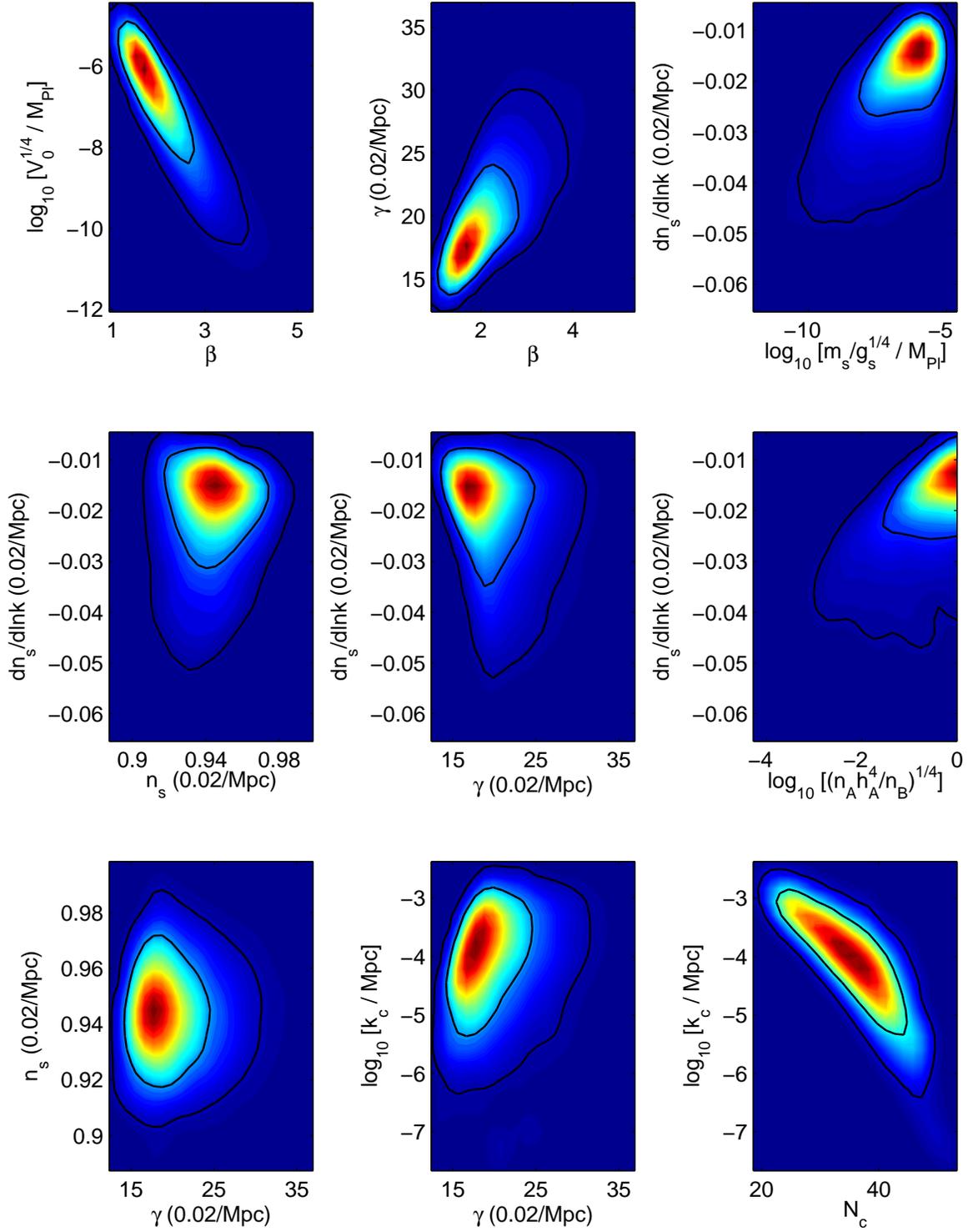}
\caption{\label{Fig:2d} 
\small Examples of 2D contours.
Solid lines show the marginalized 2D-joint 68\% and 95\% probability contours for observables and derived quantities of interest. The color coding shows the marginalized probability density in these 2D parameter spaces, ranging from red for the highest density to blue for the lowest.}
\end{center}
\end{figure}

\begin{figure}[th!]
\begin{center}
\includegraphics[scale=0.6]{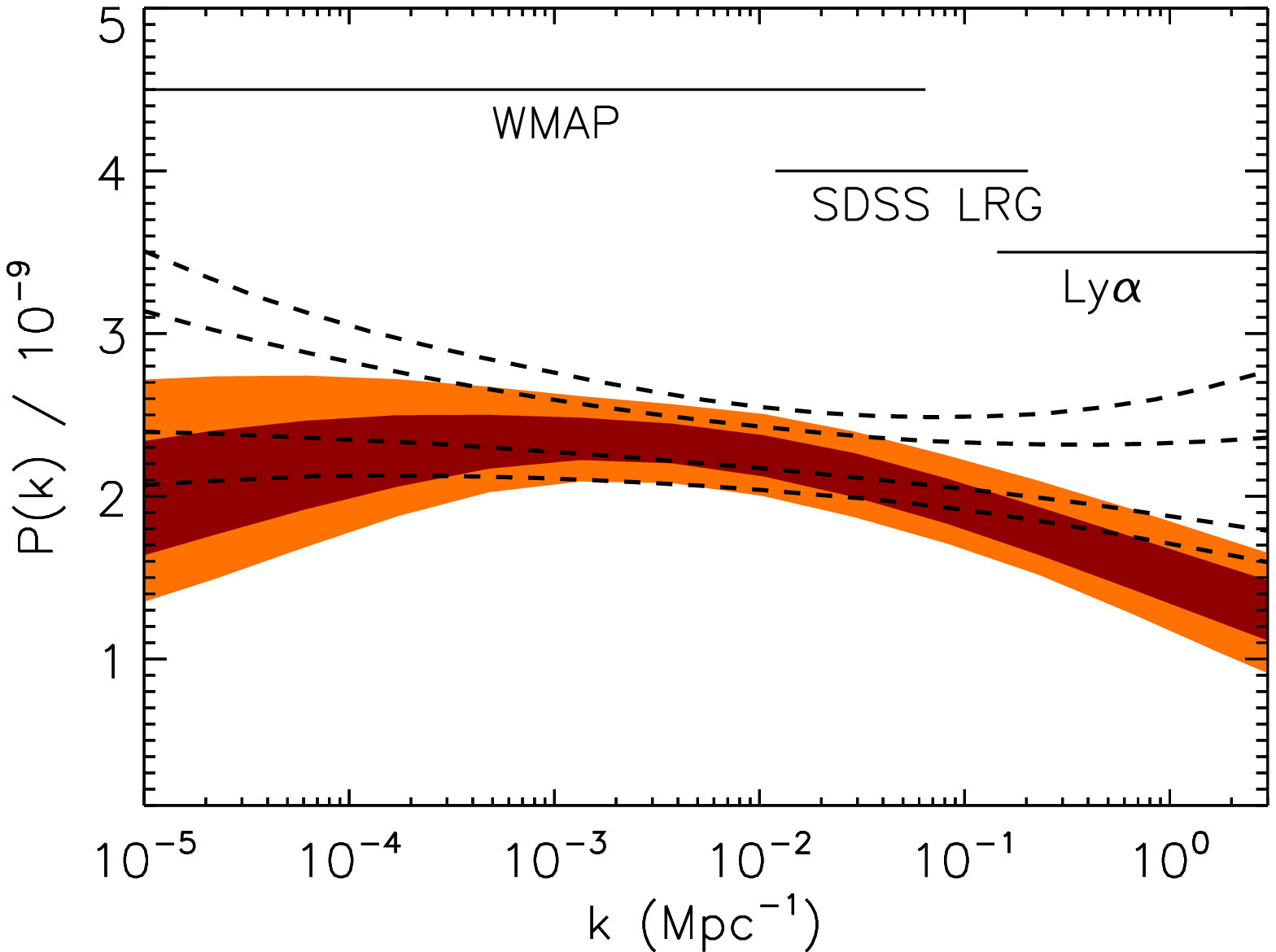} 
\includegraphics[scale=0.6]{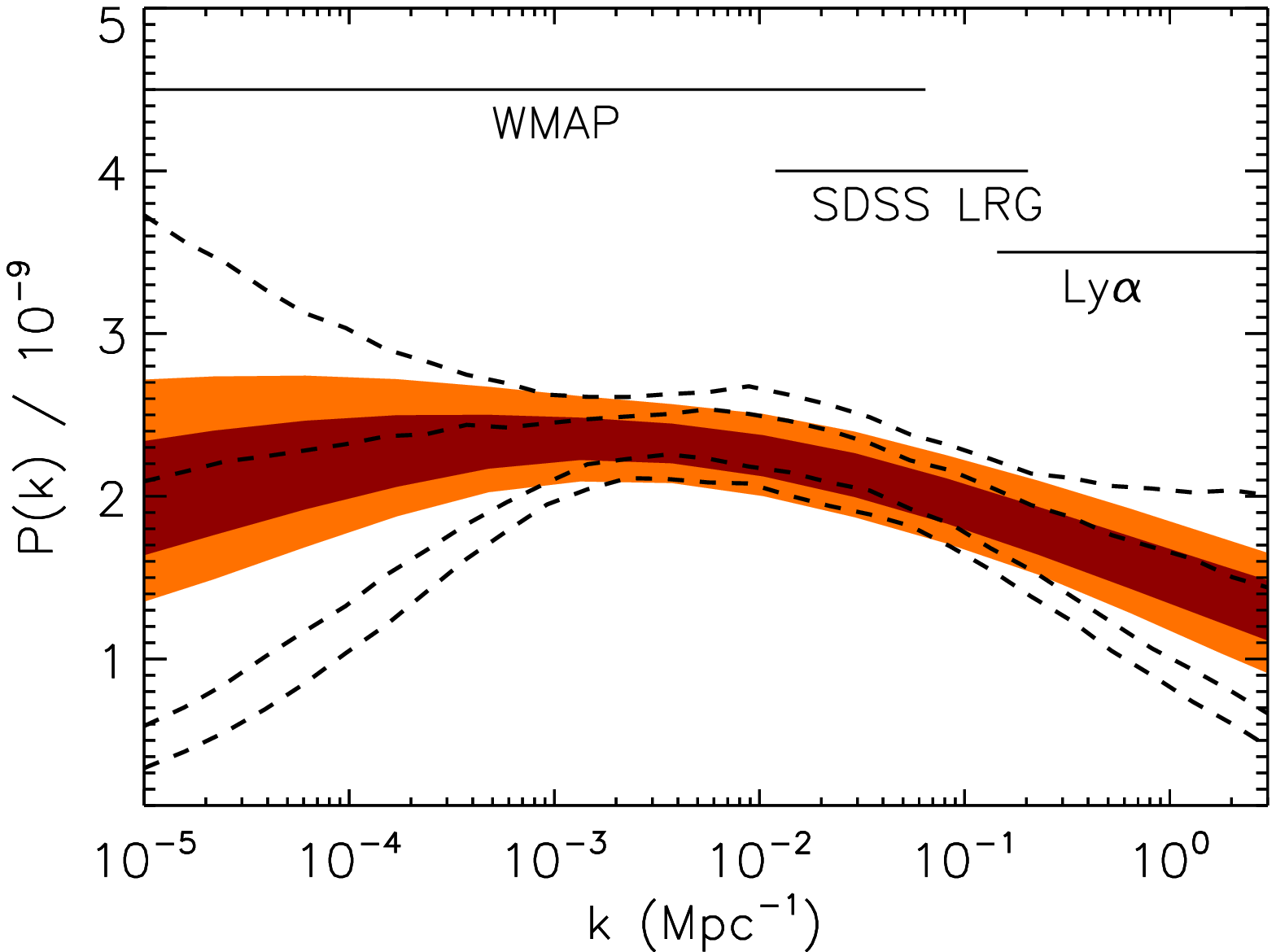} 
\end{center}
\caption{
\small Reconstructed 68\% (dark) and 95\% (light) CL constraints on
the primordial scalar power spectrum for the IR DBI model. The range
of scales spanned by WMAP and SDSS LRG data (which were used in the
fit) and the smaller scales covered by Lyman--$\alpha$ data (which
were not) are shown for reference. For comparison, the dashed lines
show the corresponding 68\% and 95\% constraints for (Upper)
single-field slow-roll inflation, taken from Ref.~\cite{Peiris:2006sj}
fitted to the WMAP and SDSS main galaxy sample data \cite{Tegmark:2003uf}, and
for (Lower) the empirical power law ansatz where the primordial power
spectrum is described by its amplitude at a pivot scale, the spectral
index $n_s$, and its running $d n_s/d \ln k$, fitted to WMAP, SDSS LRG
and Supernova Legacy Survey \cite{Astier:2005qq} data. See text for
discussion.}
\label{Fig:Pk}
\end{figure}

\begin{figure}[t!]
\begin{center}
\includegraphics[scale=0.6]{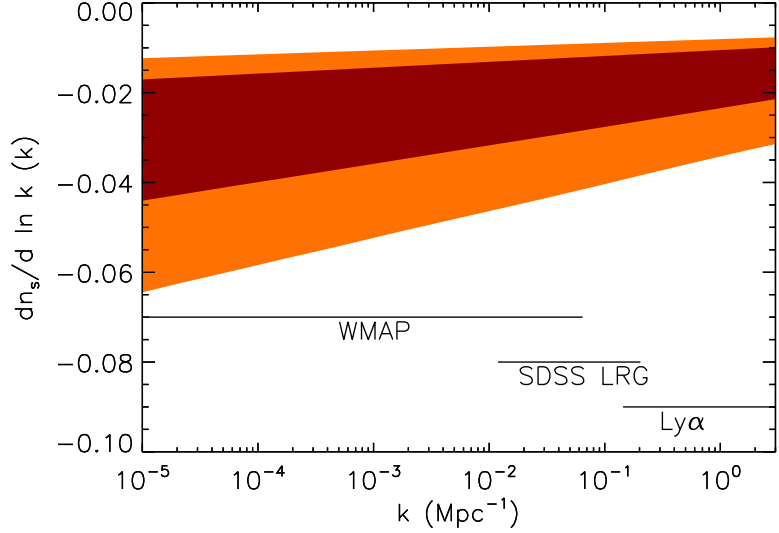}
\end{center}
\caption{
\small Reconstructed 68\% (dark) and 95\% (light) CL constraints on the scale-dependence of the running of the spectral index, $d n_s/d \ln k$, showing a mild indication for a ``running of the running''. See text for discussion.}
\label{Fig:nrun}
\end{figure}

\begin{figure}[h!]
\begin{center}
\includegraphics[scale=0.6]{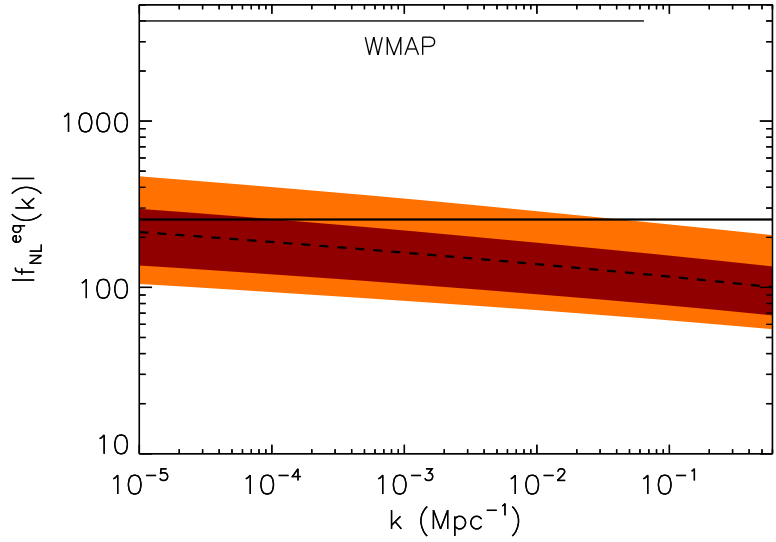}
\end{center}
\caption{
\small Reconstructed 68\% (dark) and 95\% (light) CL constraints and
mean (dashed) for the non-linearity parameter $f_{NL}^{\rm eq}$ for
the IR DBI model, as a function of scale. Note that DBI inflation 
predicts $f_{NL}^{\rm eq}<0$, and the absolute value is plotted for
convenience. The range of scales spanned by WMAP data are shown, as
well as the WMAP 95\% \emph{lower limit} $f_{NL}^{\rm eq} >-256$
(solid line), from
an analysis which treated $f_{NL}^{\rm eq}$ as scale-independent
\cite{Creminelli:2006rz}. See text for discussion.} 
\label{Fig:fnl}
\end{figure}

Following this process, we would like to apply the observational
non-Gaussianity bound $-256 < f_{NL}^{\rm eq} < 332$ (95\% CL)
\cite{Creminelli:2006rz}, as this should have a significant effect on
restricting the allowed parameter range for $\beta$ and other
parameters which are correlated with it. This is because $\gamma$ is
roughly proportional to $\beta$, and $f_{NL}^{\rm eq} = -0.32
\gamma^2$. However, two approximations enter in applying this
constraint. First, the observational constraint was obtained using an
estimator that does not encode the specific scale-dependence of the IR
DBI model, and it also does not restrict $f_{NL}^{\rm eq}<0$. Second,
Ref. \cite{Creminelli:2006rz} only gives a 95\% confidence level of
the result, and hence a full Bayesian posterior is not available for
this parameter. In order to make use of this constraint despite these
limitations, firstly we assume that the constraint of
Ref. \cite{Creminelli:2006rz} is the effective constraint at $k=0.02$
Mpc$^{-1}$, which is approximately the best constrained scale with the
current data compilation \cite{Peiris:2006sj,
Cortes:2007ak}. Secondly, we choose a Gaussian prior on $f_{NL}^{\rm
eq} (0.02\ {\rm Mpc}^{-1})$ which has the 95\% CL range found by
Ref. \cite{Creminelli:2006rz}, since the maximally uninformative prior
in the case that only a single confidence range is available has a
Gaussian form \cite{jaynes1, jaynes2}. We apply this prior to the
chains, verifying that the convergence criteria still remain
satisfied.
Finally, we obtain parameter constraints on the microphysical parameters,
cosmological parameters, cosmological observables, and derived model
parameters, which we present in
Tables~\ref{Tab_chisq}--\ref{Tab_cosmo} and
Figs.~\ref{Fig:fund_tri}--\ref{Fig:fnl}.

\begin{table}
\begin{center}
\begin{tabular}{|c|c|}
\hline 
Model & Best fit $-2 \ln\cal{L}_\mathrm{max}$ (WMAP+SDSS LRG) \\ 
\hline \hline
$\Lambda$CDM   &             5374.04 \\
 \hline
IR DBI    &    5373.11\\
 \hline
\end{tabular}
\end{center}
\caption{\small The best fit chi square, defined as $\chi^2 = -2
\ln\cal{L}_\mathrm{max}$ (where $\cal{L}_\mathrm{max}$ is the maximum
likelihood with respect to the WMAP 3 year data and the SDSS LRG
galaxy power spectrum data) for the standard $\Lambda$CDM model and
the IR DBI scenario analyzed in this work. A Gaussian prior on $f_{NL}^{\rm eq}$ has been applied based on the WMAP 3 year constraint on this parameter from Ref.~\cite{Creminelli:2006rz}. For the $\Lambda$CDM model, the $f_{NL}$ constraint has been applied assuming $f_{NL}^{\rm eq} = 0$. The IR DBI model gives a slightly better (lower) $\chi^2$ for this dataset than the
$\Lambda$CDM model. The primordial power spectrum is described by five
microphysical parameters in the former, and two empirical parameters
(an amplitude and a power law index) in the latter. When we consider
that the IR DBI observables are described phenomenologically by the
three effective parameters $N_c$, $\ln k_c$, and $n_B \beta^4$, to
which the microphysical parameters are related, we can see that the IR
DBI model has roughly one extra degree of freedom over the
$\Lambda$CDM model, which one expects to give a  $\Delta \chi^2 \sim
1$ improvement in the fit. Since this is in fact what we see, there is
no indication of a preference in the data for the IR DBI model.}
\label{Tab_chisq}
\end{table}

\begin{table}
\begin{center}
\begin{tabular}{|c|c|c|}
\hline 
Parameter & Marginalized Constraint & Maximum Likelihood \\ 
\hline \hline 
$\Omega_b h^2$ & $0.02145_{-0.00071-0.00138}^{+0.00071+0.00138}$  &  0.02162 \\
\hline
$\Omega_c h^2$ &  $0.1070_{-0.0044-0.0082}^{+0.0042+0.0086}$ & 0.1058  \\
\hline
$\tau$ &  $0.089_{-0.030-0.061}^{+0.030+0.060}$ &  0.094 \\
\hline
$H_0$ & $71.2_{-1.9-3.7}^{+1.8-3.9}$  &  72.1 \\
\hline
\hline
$\log_{10} [n_B]$ & $4.64_{-0.32-0.70}^{+0.30+0.45}$ &  4.93 \\
\hline
$\log_{10} [m_s/g_s^{1/4} / M_{\rm Pl}]$ &  $-6.71_{-1.07-2.89}^{+1.04+1.43}$ &  $-5.91$ \\
\hline
$\beta$ & $2.11_{-0.60-0.85}^{+0.63+1.63}$ &  1.77 \\
\hline
$\log_{10} [N_B]$ & $9.48_{-0.39-0.70}^{+0.39+0.93}$  & 9.15 \\
\hline
$\log_{10} [n_A h_A^4]$ & $1.41_{-2.82-8.02}^{+2.64+3.34}$  &  $0.585$ \\
\hline
\hline
$\log_{10} [V_0^{1/4} / M_{\rm Pl}]$ &
$-6.95_{-1.34-3.09}^{+1.25+1.83}$ & $-6.36$  \\
\hline
$N_c$ & $35.7_{-7.3-12.6}^{+6.8+11.7}$ & 34.1  \\
\hline
$\log_{10}$ [k$_c$/Mpc] & $-4.15_{-0.81-1.82}^{+0.81+1.21}$ & $-3.86$    \\
\hline
$N_{\rm tot}^{\rm NR}$ & $18.4_{-3.2-5.7}^{+3.3+5.8}$ &  20.5  \\
\hline
$N_e^{\rm DBI}$ (10$^{-5}$/Mpc) & $38.4_{-6.1-10.7}^{+5.6+9.1}$ & 37.6  \\
\hline
$\log_{10} [(n_Ah_A^4/n_B)^{1/4}]$ & $> -2.36$ (95\% CL) &   $-1.09$ \\
\hline
$\log_{10} [(h_A m_s/g_s^{1/4})^2 / (16\pi^2)/M_{\rm Pl}^2]$ &
$-17.2_{-2.6-6.0}^{+2.4+3.5}$  &  $-16.2$ \\
\hline
$n_s$ (0.02/Mpc) & $0.943_{-0.016-0.031}^{+ 0.016+0.032}$ & 0.946  \\
\hline
$dn_s/d\ln k$ (0.02/Mpc) & $-0.021_{-0.009-0.025}^{+0.008+0.011}$  & $-0.021$  \\
\hline
$\gamma$ (0.02/Mpc) & $19.9_{-3.4-5.1}^{+3.6+9.3}$   &  16.8  \\
\hline
$f^{\rm eq}_{NL}$ (0.02/Mpc) & $-131_{-45-141}^{+44+61}$ & $-91$ \\
\hline
\end{tabular}
\end{center}
\caption{\small
Constraints on the IR DBI model from the WMAP and SDSS LRG data-sets
(mean, upper and lower 68\% and 95\% CL, marginalizing over all other
parameters), and the maximum likelihood values of the parameters found
in the MCMC. A Gaussian prior on $f_{NL}^{\rm eq}$ has been applied based on the WMAP 3 year constraint on this parameter from Ref.~\cite{Creminelli:2006rz}.}
\label{Tab_cosmo}
\end{table}

\clearpage

\vspace*{2cm}
\clearpage

\clearpage

\clearpage

\section{Discussion and Conclusions}\label{conc}

We conclude by highlighting the main results and discussing their
physical implications.
The quoted ranges are at the $95\%$ confidence level, 
and we have combined
constraints from both the power spectrum and non-Gaussianity.
The detailed $68\%$ and $95\%$ CL marginalized constraints and
the maximum likelihood values are listed in Table \ref{Tab_cosmo}.

\subsection{Microscopic parameters}

\begin{itemize}

\item {\em Shape of the inflaton brane moduli potential:
$1.3<\beta < 3.7$.}
The lower bound is due to constraints from the power spectrum, while
the upper bound is due to the non-Gaussianity constraint.
It is encouraging that, while IR DBI inflation can happen for a range
of $\beta$ that varies over nearly 10 orders of magnitude, $0.1
\lesssim\beta <10^9$ (see Eq.~(\ref{mrange})), comparison with data
picks out a very small range around $\CO(1)$ which is generically
expected theoretically. 
This makes an explicit construction of such potentials a more
interesting question.

\item {\em Fundamental string scale: $-9.6 <
\log_{10}(m_s/\mpl)/g_s^{1/4} < -5.3 $.}
The upper bound on the
string scale is due to the large charge, and hence length scale, of
the B-throat required to fit the amplitude of the density
perturbations. The lower bound is due to the fact that a smaller string
scale tends to increase the total number of $e$-folds of non-relativistic
fast-roll inflation, and make the running of the spectral index too
large (Fig.~\ref{Fig:2d}).
The model prefers an intermediate fundamental string scale, $10^8\ {\rm
GeV} < m_s/g_s^{1/4} < 10^{13}\ {\rm GeV}$, and therefore an
intermediate 
large volume compactification, $8.9\times 10^7 < V^{1/6} \mpl
<4.8\times 10^{13}$,
where $V$ is the compactification volume.

\item {\em B-throat charge: $8.8< \log_{10} N_B < 10.4 $; 
Number of inflaton branes: $3.9<\log_{10} n_B< 5.1$.}
In terms of the GKP-type warped compactification, this implies flux
numbers $K\sim M \sim \sqrt{N_B} \sim \CO(10^5)$. 
Explicit construction
remains an open question as discussed in Sec.~\ref{Sec:OpenQu}.
In the multi-throat brane inflation scenario, 
inflaton branes are generated from flux-antibrane
annihilation. The number of branes generated in this process is
roughly determined by the flux number $M$. Indeed, a small number of
inflaton branes is ruled out by the data.

\item {\em A-throat minimum warp factor: $-2.4<\log_{10} h_A \le 0$.} 
This is
from combining the constraint on $n_B$ and $n_A h_A^4$,
$h_A = (n_A h_A^4/n_B)^{1/4}$.
A smaller $h_A$ leads to larger $N^{\rm NR}_{\rm tot}$ and larger
running of the spectral index (Fig.~\ref{Fig:2d}).
So the A-throat tends to be short. This makes  
tunneling reheating possible, where many interesting phenomena
can occur, such as an intermediate matter-dominated epoch.

\end{itemize}

\subsection{Secondary derived parameters}

\begin{itemize}

\item {\em Inflationary phases.}
In this model, not all $e$-folds comes from IR DBI inflation. The
last $13< N^{\rm NR}_{\rm tot}< 24$ $e$-folds
come from non-relativistic fast-rolling
inflation, which is possible because inflatons are close to the top of
the potential.

\item {\em The stringy phase transition.}
The Hubble-expansion induced stringy phase transition 
happens at the largest scales in the sky, 
$-6.0<\log_{10} k_c/{\rm Mpc} <-2.9$. 
However its impact on density perturbations extends over to
shorter scales, such as generating a transient large
running of the spectral index.

\item {\em Inflation scale:
$-10.0<\log_{10}V_0^{1/4}/\mpl<-5.1$.}
This gives a very small tensor to scalar ratio $r_{TS}<10^{-13}$.

\item {\em Cosmic string tension: $-23<\log_{10} G\mu_D 
+\log_{10}g_s^{1/2} < -14$.}
Here the cosmic strings refer to
the D-strings left over from the brane-antibrane annihilation in the
A-throat, whose tension is
$G\mu_D= (m_s h_A/g_s^{1/4} \mpl)^2/(16\pi^2 g_s^{1/2})$.
There is an unconstrained freedom coming from the additive factor
$\log_{10}g_s^{1/2}$, but it is not expected to give any significant
contributions.
The F-string tension differs by a factor of $g_s$,
$\mu_F = g_s \mu_D$.

\end{itemize}

\subsection{Observational predictions}

\begin{itemize}

\item {\em Large, but regional, running of spectral index:
$-0.046< dn_s/d\ln k(k=0.02/{\rm Mpc}) <-0.010$.}
A reconstructed full-scale power spectrum and the running of the
spectral index are shown in Fig.~\ref{Fig:Pk} \& \ref{Fig:nrun}.

This prediction is stringy in nature. A better understanding of
the theoretical details and better measurements of both the power spectrum
and non-Gaussianities on the relevant scales
may reveal finer structures. In future experiments, 
Planck is expected to achieve
$\sigma(dn_s/d\ln k) = 0.005$ \cite{Planck}.

\item {\em Large non-Gaussianities: 
$-272<f_{NL}^{\rm eq}(k=0.02/{\rm Mpc}) <-70$.}
A reconstructed full-scale prediction is in Fig.~\ref{Fig:fnl}, 
which shows the running of the non-Gaussianities.

This prediction is strictly speaking field-theoretic, but with strong
string theory motivations, such as warped compactification and the DBI
brane action. 
This field theoretic regime is $k>k_c$; the theoretical analysis for
non-Gaussianities at $k\lesssim k_c$ is currently unavailable and
remains an interesting open question.
In future experiments,
on CMB scales, Planck can achieve $\sigma(f^{\rm
eq}_{NL}) =67$ \cite{Smith:2006ud,Hikage:2006fe}; 
on large scale structure scales, some high-$z$ galaxy surveys can reach
similar or better precision \cite{Sefusatti:2007ih}.

\end{itemize}

As seen from these results, constraints from cosmological data, and
even relatively loose constraints such as the non-Gaussianity
constraint, are already putting strong restrictions on models which
aim to provide self-consistent microphysical descriptions of the early
universe. With the bounty of precision cosmological data expected in
the future, the hope of probing not just field-theoretic, but
string-theoretic early universe physics burns brightly.

\medskip
\section*{Acknowledgments}

We thank Alan Guth, Shamit Kachru,
Hong Liu, Marilena LoVerde, Andrew Lynch, Daniel Mortlock, Sarah
Shandera, Eva Silverstein, and
Henry Tye for helpful discussions in the course of this
work. We acknowledge use of the Legacy Archive for Microwave
Background Data Analysis (LAMBDA). 
XC thanks the hospitality of the organizers of the ``Life beyond the
Gaussian'' workshop at the KICP at the University of Chicago
where part of the work was initiated.
This work was partially supported
by the National Center for Supercomputing Applications under grant
number TG-AST070004 (HVP, PI) and utilized computational resources on the
TeraGrid (Cobalt). RB's work is supported by NSF grants AST-0607018
and PHY-0555216. XC is supported by 
the US Department of Energy under cooperative research agreement
DEFG02-05ER41360 and
the National Science Foundation under grant PHY-0355005.
HVP is supported by NASA through Hubble Fellowship
grant \#HF-01177.01-A from the Space Telescope Science Institute,
which is operated by the Association of Universities for Research in
Astronomy, Inc., for NASA, under contract NAS 5-26555, and by a STFC
Advanced Fellowship.
JX is supported in part by 
the National Science Foundation under grant PHY-0355005.


\appendix

\section{Estimate the effect of phase transition on spectral index}
\label{Sec:Width}
\setcounter{equation}{0}

In this appendix we estimate the transition behavior of the spectral
index between two asymptotic values described in
Ref.~\cite{Chen:2005ad,Chen:2005fe} and Sec.~\ref{Sec:Power}. Consider
a simple model where the density perturbations are caused by the
scalar field fluctuations, which are the super-horizon  ripples on
branes in transverse directions. These ripples are generated
during a Hubble time while they are still sub-horizon and then
frozen. The amplitude of the ripples are given by the fluctuation
speed of a Hubble-sized patch on the brane. This speed is determined
by the energy pumped into the branes by the Hubble expansion. This
model simplifies the underlying physics by focusing on only 
the overall fluctuation speed of a Hubble-sized patch while ignoring
the detailed world-volume theory such as effects from 
specific stringy excitations.

According to the special relativity, an object with rest mass $m_0$ and
energy $E=m_0 + \Delta E$ has velocity
\bea
v= c \sqrt{1- \frac{m_0^2}{(m_0+\Delta E)^2}} ~.
\eea
For the on-brane observer, a Hubble-sized patch has rest mass 
\bea
m_0 = h_B^4 T_3 \Delta V = h_B^4 T_3 \left( \frac{\gamma H}{2\pi}
\right)^{-3} ~, 
\eea
where $h_B^4 T_3$ is the red-shifted brane tension. The Hubble energy
is $\gamma H/2\pi$, half
of which goes to the kinetic energy of the transverse oscillation of
the brane $\Delta E = \gamma H/4\pi$, while the other half goes to the
tension of oscillations in terms of spatial derivatives. We have
restored the factor of $2\pi$ in the Hubble length and energy in order
to quantitatively match the known results in the low energy limit. The
local speed of light is $c=h_B^2$. The position-dependent time delay
is 
\bea
\delta t = \frac{v (\gamma H/2\pi)^{-1}/\gamma}{ \dot r \sqrt{n_B}} ~,
\eea
where the numerator is the fluctuation amplitude within a Hubble time
viewed from the lab observer (hence an extra factor of $1/\gamma$ due
to Lorentz contraction), and the denominator is the overall brane
velocity which is approximately the local speed of light $\dot r
\approx c$. Here we also consider the case of $n_B$ multiple branes
where the superposition of
independent fluctuations reduces the time-delay by a factor of
$1/\sqrt{n_B}$.
Using these estimates we obtain the power spectrum
\bea
P_k = \frac{4\pi^2 v^2 T_3}{\gamma^4 \dot \phi^2} ~,
\label{PkallA}
\eea
where
\bea
v^2 T_3 = h_B^4 T_3 \left[1- \left(1+
\high{ \frac{\gamma^4
H^4}{32\pi^4 h_B^4 T_3} }\right)^{-2} \right] ~.
\label{v2T3A}
\eea
This formula recovers the usual field theory result in the limit of non-relativistic fluctuation speed. This includes non-relativistic-(slow or fast)-roll inflation, and DBI inflation below the phase transition. 
This formula also gives an estimate on the effect of the
Hubble-expansion
induced stringy phase transition. The estimate is expected to provide the envelope behavior beyond the transition since it ignores detailed
features such as specific resonant production of various stringy states.

It will be useful to extract the DBI inflation region in
(\ref{PkallA}) and (\ref{v2T3A}), and parametrize it in the following
way,\footnote{More precisely, because of the sound horizon is
time-dependent, we should replace $N_e^{\rm DBI}$ in (\ref{PkallDBI})
with $N_e^{\rm DBI}-\ln (c_s(k) H(k)^{-1})/(\hat c_s \hat H^{-1})$,
where the variables with a hat are evaluated when the reference mode
$\hat k$ (e.g.~$\hat k= 0.002/Mpc$ as in (\ref{Nek})) crosses the
sound horizon. Because the relevant scales for (\ref{PkallDBI}) span
only a few $e$-folds, the change of the sound horizon $c_s H^{-1}$ is
very small and we neglect such corrections.}
\bea
P_k = H^2 \delta t^2 = \frac{324\pi^2}{n_B \beta^4 {N_e^{\rm DBI}}^4}
\left( 1- \frac{N_c^{16}}{(N_c^8 + {N_e^{\rm DBI}}^8)^2} \right) ~.
\label{PkallDBI}
\eea
$N_c$ is defined as
\bea
N_c \equiv 2^{5/8} \sqrt{3\pi} \frac{\lambda_B^{1/8}}{n_B^{1/8}
\beta^{1/2}} = \sqrt{6}\pi^{1/4}
\frac{ N_B^{1/8}}{\beta^{1/2}} ~,
\label{Ncdef}
\eea
where we have used the relation (\ref{lambdaDef}).
Taking the limits $N_e \ll N_c$ and $N_e \gg N_c$, we recover (\ref{PkDBI}) and (\ref{Pklarge}) respectively.
The spectral index is
\bea
n_s-1 = \frac{4}{N_e^{\rm DBI}} \frac{x^2+3x-2}{(x+1)(x+2)} ~, 
~~~~ x\equiv \left(\frac{N_e^{\rm DBI}}{N_c} \right)^8 ~.
\label{nsallA}
\eea
This formula interpolates between two asymptotic values 
$4/N_e$ and $-4/N_e$. If we define the width of the transition region
as the $e$-fold difference $\Delta N_e$ between $n_s-1 = 2/N_e^{\rm DBI}$ 
and $-2/N_e^{\rm DBI}$, then we have 
\bea
\Delta N_e \approx 0.2 N_c ~,
\label{DeltaNc}
\eea
which can be quite large (for example, six if $N_c = 30$).
But the running of $n_s$ is still observably 
large in the transition region 
(for example, $dn_s/d\ln k \approx -0.02$ in the range of
(\ref{DeltaNc}) for $N_c=30$).

\section{Running spectral index from slow-roll potential with mild
features}
\label{Sec:Mild}
\setcounter{equation}{0}

Usual slow-roll inflation gives negligible running of the spectral
index, $dn_s/d\ln k = \CO(10/N_e^2)$, 
because large running of $n_s$ tends to end the
inflation too quickly.
For a comparison with data, see Ref.~\cite{Easther:2006tv}.
In this appendix, we study the possibility of
having measurable $|dn_s/d\ln k| \gtrsim 0.01$ 
by adding some mild features to the slow-roll potential, and 
how we can phenomenologically distinguish it from 
the IR DBI inflation model.

We consider a potential of a small field inflation and add some
ripples on it,
\bea
V= V_0 -a \phi -b \sin (\phi/\phi_0) ~.
\label{Vripples}
\eea
The inflaton starts for example at $\phi_i=0$.
At $\phi=\phi_{\rm end}$, one imagines that the
inflationary energy $V_0$ gets annihilated as in brane-anti-brane
inflation.
This is just an example of many possibilities, which we use to
illustrate the main properties. As we will see,
to generate a {\em large} running of $n_s$ from blue to red, the
shape of the 
slow-roll potential changes from convex to concave. The
oscillatory ripples help to sustain inflation, and at the same
time generate large $dn_s/d\ln k$ periodically.
In fact, for our purpose, it is not necessary to make the mild feature
periodic, for example the 3rd term in (\ref{Vripples}) can be regional
as long as it falls into the
WMAP range. Nonetheless being periodic might be more naturally 
realized in model-building.

We want the inflaton to continuously roll down, so we need
$V'\le 0$, {\it{i.e.}}~, 
\bea
-a + b/\phi_0 \le 0 ~.
\label{Vslope}
\eea
We require the average slow-roll parameter
\bea
\epsilon = \frac{\mpl^2}{2} \frac{a^2}{V_0^2} \ll 1 
\eea
to have enough inflationary $e$-folds.
To have the effect of one ripple span several $e$-folds, we need
$\phi_0/\dot \phi = \xi H^{-1}$, where $\xi$ is of order one.
$\dot\phi$ can be estimated
using the attractor behavior $3H\dot \phi + V' =0$ and taking the
average value of $V'$.
We want the other slow-roll parameter $\eta$ to vary between order
$\pm 0.1$ to
generate observable $dn_s/d\ln k$,
\bea
|\eta| \le \mpl^2 \frac{b}{\phi_0^2 V_0} \equiv \zeta ~,
\eea
where $\zeta$ is $\CO(0.1)$.

\begin{figure}[p]
\begin{center}
\includegraphics[scale=0.6]{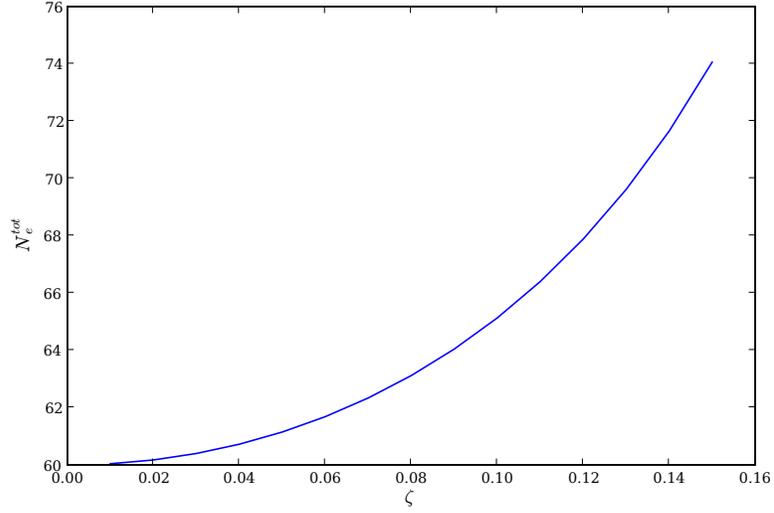}
\end{center}
\caption{\small Turning on the periodic mild features $\zeta$ does
not significantly affect the total inflationary $e$-folds. In this
figure, $\xi=4$, $a/V_0=10^{-4}$, $\phi$ starts from $0$ and ends at
$0.006\mpl$.}
\label{Fig:srmild_Ne}
\end{figure}

\begin{figure}[p]
\begin{center}
\includegraphics[scale=0.75]{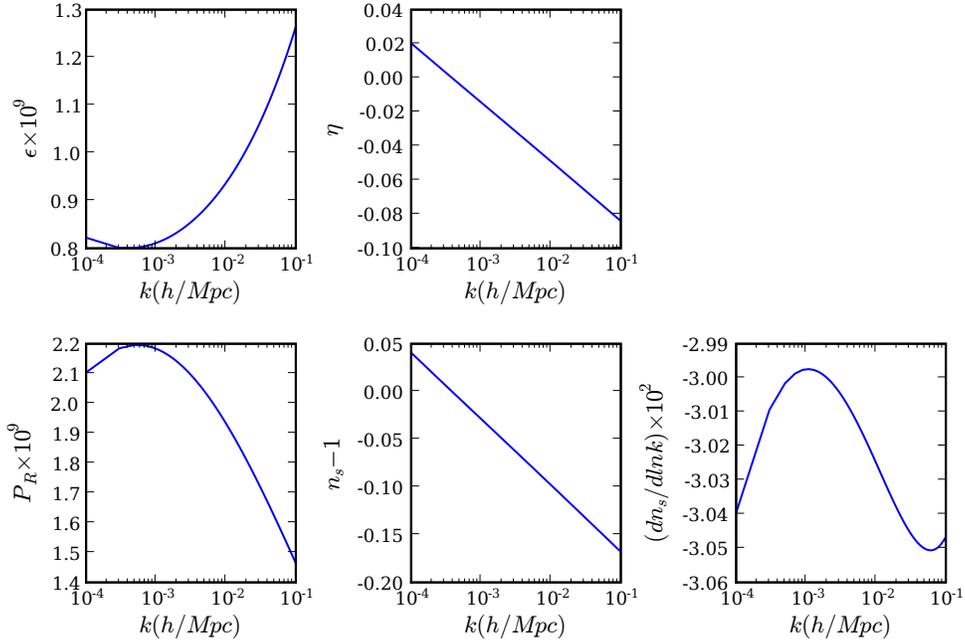}
\end{center}
\caption{\small The power spectrum and spectral index in the WMAP
range for a slow-roll
potential with mild
features. The parameters $a/V_0=10^{-4}$, $\xi=4$ and $\zeta = 0.15$
are chosen so that these observables look close to what we obtained 
in the IR DBI model.}
\label{Fig:srmild_obvs}
\end{figure}

Therefore for our purpose, we can choose the parameters in
(\ref{Vripples}) in the following way. We require
\bea
a \ll \frac{V_0}{\mpl} 
\eea
so that $\epsilon \ll 1$; 
\bea
\phi_0 = \xi \frac{a\mpl^2}{V_0}
\eea
with $\xi$ of $\CO(1)$
so that the effect of one period of the mild feature spans a reasonable
amount of $e$-folds;
\bea
b = \xi^2 \zeta \frac{a^2 \mpl^2}{V_0} 
\eea
with $\zeta$ of $\CO(0.1)$
so that the running of $n_s$ is observably large.
In order for these conditions to be consistent with (\ref{Vslope}), we
need
\bea
\xi \zeta \le 1 ~.
\eea
For example, we can choose $\xi=4$, $\zeta = 0.15$.

In Fig.~\ref{Fig:srmild_Ne} and \ref{Fig:srmild_obvs} we demonstrate
numerically:
\begin{itemize}

\item Introducing such mild features does not significantly affect, and in
fact can slightly increase, the
total number of inflationary $e$-folds.

\item The power spectrum within the WMAP scales looks like
what we obtained in IR DBI model; notably, the spectral index runs from
blue to red with a large and negative $dn_s/d\ln k$.

\end{itemize}

To experimentally distinguish this case from the IR DBI model, 
we estimate the non-Gaussianity. In slow-roll inflation,
the 3-point function of the gauge-invariant scalar perturbation
receives contributions from the following sources
\cite{Maldacena:2002vr,Seery:2005wm}. In the cubic
action there are terms proportional to $\epsilon^2$, $\epsilon^3$ and
$\epsilon d\eta/dt$. 
In terms of order-of-magnitude, these terms
contribute $\CO(\epsilon)$, $\CO(\epsilon^2)$, $\CO(\Delta \eta)$, 
respectively, to the non-Gaussianity estimator $f_{NL}$. The $\Delta
\eta$ is the maximum change of $\eta$ caused by the features, since the
3-point function involves an integration over time.
In addition, there is a field redefinition term that contributes
$\CO(\eta_{\rm end})$ to $f_{NL}$, where $\eta_{\rm end}$ is the frozen
value of $\eta$ after the horizon crossing. 

In case of slow-roll inflation with smooth potential, the leading
terms of $f_{NL}$ are $\CO(\epsilon)$ and $\CO(\eta)$
\cite{Maldacena:2002vr}. 
In case of sharp
features, $\CO(\Delta \eta)$ term dominates \cite{Chen:2006xj}. 
In the case of interest
here with periodic mild features, both $\CO(\Delta \eta)$ and the
boundary term $\CO(\eta_{\rm end})$ become important.
As we saw, to generate large but reasonable running of the spectral
index, we require $\eta$ to vary between $\pm\CO(0.1)$. 
So we expect such features to
be associated with non-Gaussianities $f_{NL} = \CO(0.1)$. This is
clearly observationally distinguishable from the IR DBI
inflation case.

\section{Running spectral index from non-Bunch-Davies vacuum}
\label{Sec:nonBD}
\setcounter{equation}{0}

In this Appendix, we study how the running spectral index arising from
the non-Bunch-Davies vacuum case
can be phenomenologically distinguished from the running spectral
index in IR DBI model.

In the field theory of density perturbations, the Bunch-Davies (BD) 
vacuum
is the leading behavior of the fluctuations when they are well within
the horizon. However corrections to such a vacuum can have
observational effects and may provide information on new physics
\cite{Martin:2000xs}.
For reviews and references see
\cite{Brandenberger:2005be,Greene:2005aj}. 
Denoting the scale of the new
physics as $M\gg H$, this correction typically arises at the order
$H/M$,
for example, if we choose the adiabatic vacuum at the scale $M$
\cite{Danielsson:2002kx}. This
is also called the trans-Planckian effect if $M$ is regarded as
$\mpl$.
In this appendix, we will treat $M$ to be much more general.

The main difference between the case of non-BD vacuum 
and the Hubble-expansion induced stringy
phase transition in DBI inflation is that, 
in the former, there is a large region
between the new physics scale and the Hubble horizon where the
conventional field theoretic analyses still 
applies; while in the latter, the
Hubble horizon is comparable to or smaller than the new stringy length
scale in the inflaton sector.

The effect of non-Bunch-Davies vacuum in slow-roll inflation
typically results in an
oscillatory modulation
on the usual power spectrum \cite{Easther:2001fz,Easther:2002xe}. 
So it may also introduce an observable
running spectral index. The potential for observing these features are
discussed in Ref.~\cite{Easther:2004vq,Easther:2005yr}.
To study both slow-roll and DBI inflation, here we generalize the
analyses of Ref.~\cite{Danielsson:2002kx,Polarski:1995jg} 
to the case with arbitrary sound speed.

The quadratic action for the gauge-invariant scalar perturbation
$\zeta$ in general single field inflation is 
\bea
S_2  = \int dt d^3x \left[ a^3 \frac{\epsilon}{c_s^2} \dot \zeta^2 - a
\epsilon (\partial \zeta)^2 \right] ~,
\eea
where $a$ is the scale factor, $c_s$ is the sound speed, and
$\epsilon = -\dot H/H^2$ is one of the slow variation parameters.
Using the variable $v_\bk\equiv z \zeta_\bk$ 
($z\equiv a\sqrt{2\epsilon}/c_s$) and 
its conjugate momentum 
\bea
\pi_\bk = v'_\bk - \frac{z'}{z} v_\bk ~,
\eea
the Hamiltonian in Fourier space is 
\bea
H_2 = \frac{1}{(2\pi)^3} \int d^3 k \half 
\left[ \pi_\bk \pi^*_\bk + k^2 c_s^2 v_\bk v^*_\bk
+ \frac{z'}{z} ( \pi_\bk v^*_\bk + \pi^*_\bk v_\bk) \right] ~,
\eea
where the prime denotes the derivative with respective to the
conformal time $\tau$.

We can quantize $v_\bk$ and $\pi_\bk$
in terms of the creation and annihilation operators which are either
time-dependent
\bea
v_\bk &=& \frac{1}{\sqrt{2k c_s}} \left( a_\bk (\tau) +
a^\dagger_{-\bk}(\tau) \right) ~, \nonumber \\
\pi_\bk &=& -i \sqrt{\frac{k c_s}{2}} \left( a_\bk(\tau) - 
a^\dagger_{-\bk}(\tau) \right) ~,
\label{Hpic}
\eea
or time-independent
\bea
v_\bk &=& f_k(\tau) a_\bk(\tau_0) + f_k^*(\tau)
a^\dagger_{-\bk}(\tau_0) ~, \nonumber \\
\pi_\bk &=& -i \left[ g_k(\tau) a_\bk(\tau_0)
-g^*_k(\tau) a^\dagger_{-\bk}(\tau_0) \right] ~,
\label{Spic}
\eea
where $f_k(\tau)$ is the solution of the equation of motion
$v_\bk'' + c_s^2 k^2 v_\bk - (z''/z) v_\bk = 0$,
\bea
f_k(\tau) &=& C_+ \frac{1}{\sqrt{2c_s k}}
\left(1-\frac{i}{k c_s\tau}\right) e^{-ik c_s\tau}
+ C_- \frac{1}{\sqrt{2c_s k}}
\left(-1-\frac{i}{k c_s\tau}\right) e^{ik c_s\tau} ~,
\nonumber \\
g_k(\tau) &=& C_+ \sqrt{\frac{k c_s}{2}} e^{-ik c_s\tau}
+ C_- \sqrt{\frac{k c_s}{2}} e^{ik c_s\tau} ~.
\eea
Equations (\ref{Hpic}) and (\ref{Spic}) 
are related by the Bogolubov transformation
\bea
a_\bk(\tau) &=& \xi_k(\tau) a_\bk(\tau_0)
+ \zeta_k(\tau) a^\dagger_{-\bk}(\tau_0) ~, \nonumber \\
a^\dagger_{-\bk}(\tau) &=& \xi^*_k(\tau) a^\dagger_{-\bk}(\tau_0)
+ \zeta_k^*(\tau) a_\bk(\tau_0) ~,
\label{Btran}
\eea
where
\bea
\xi_k &=& \sqrt{\frac{k c_s}{2}} f_k + \sqrt{\frac{1}{2k c_s}} g_k ~,
\nonumber \\
\zeta_k &=& \sqrt{\frac{k c_s}{2}} f^*_k -
\sqrt{\frac{1}{2k c_s}} g^*_k ~.
\eea
The following relation should be satisfied to preserve the commutation
relation for (\ref{Btran}),
\bea
|\xi_k|^2 - |\zeta_k|^2 =1.
\eea
Therefore we have
$f_k g_k^* + f_k^* g_k =1$ and $|C_+|^2-|C_-|^2=1$.

An adiabatic vacuum $|0,\tau_0\rangle$ can be chosen as 
\bea
a_\bk(\tau_0) |0,\tau_0\rangle = 0 ~, ~~~~{\rm {\it{i.e.}}}~~~
\zeta_k(\tau_0)=0 ~.
\eea
The Bunch-Davies vacuum corresponds to sending $\tau_0$ to
$-\infty$. More generally one can choose a finite $\tau_0\approx
-1/a_0 H_0$ for the mode $\bk$, 
when this mode crosses the scale of the new physics $M \gg
H/c_s$. Hence the relation between the power spectrum in the
non-BD and BD vacuum is
\bea
P_k^{\rm nonBD} = |C_+ + C_-|^2 P_k^{\rm BD} ~,
\label{PknonBD}
\eea
where
\bea
|C_+ + C_-|^2 &\approx&
1- \left( \frac{a H}{k c_s} \right)_{\tau} 
\sin \left( \frac{2k c_s}{a H} \right)_{\tau}
~, \nonumber \\
&\approx&
1- \left( \frac{H}{M c_s} \right)_{\tau} 
\sin \left( \frac{2M c_s}{H} \right)_{\tau} ~,
\eea
where the new physics scale $M = (k/a)_{\tau}$.
The extra contribution of the non-BD vacuum to the spectral index is
\bea
\Delta n_s \approx
-2(\epsilon+s+\mu) \cos \frac{2M c_s}{H} ~,
\label{nsnonBD}
\eea
where $\mu \equiv \dot M/(HM)$, $s\equiv \dot c_s/(Hc_s)$.
In the following discussion, we will concentrate on the amplitude and
oscillation frequency of these features on the power spectrum and
spectral index. We want to compare them to those in IR DBI model, 
where
the spectral index $n_s-1$ changes 
between $\pm \CO(0.1)$ within $\CO(10)$
$e$-folds without oscillations.
For this purpose it is useful to note that 
the change of the arguments in
the trigonometric functions in (\ref{PknonBD}) and (\ref{nsnonBD}) as
a function of $k$ can
be written as
\bea
\Delta \left( \frac{2M c_s}{H} \right)_\tau =
\left( \frac{2M c_s}{H} \right)_{\tau_0} 
(\epsilon+s+\mu) \ln \frac{k}{k_0} ~.
\label{DeltaA}
\eea

We first look at slow-roll inflation, where $s=\mu=0$ and
$\epsilon \lesssim 0.01$. So the variation of $n_s$, typically smaller
than $\CO(0.01)$, is much
smaller than that caused by 
the phase transition in IR DBI case, although
the oscillatory frequency of the $n_s$ is adjustable
depending on the values of $(2M/H)_{\tau_0}$ and $\epsilon$.
For $n_s$ to have larger variations, one needs the special case of
$\epsilon \approx 0.05$; at the same time, for the running to span
$\CO(10)$ $e$-folds without oscillation, from (\ref{DeltaA}), we see
that $Mc_s/H = \CO(\pi)$. This barely satisfies $Mc_s/H \gg 1$.
We conclude that in slow-roll inflation 
the effect of the non-DB vacuum on $n_s$, having much
smaller $\Delta n_s$ or large oscillatory frequencies, 
will be observationally distinguishable from
that caused by the phase transition in IR DBI inflation.
In addition, there are no observable non-Gaussianities associated with
them.

\begin{figure}[t]
\begin{center}
\includegraphics[scale=0.7]{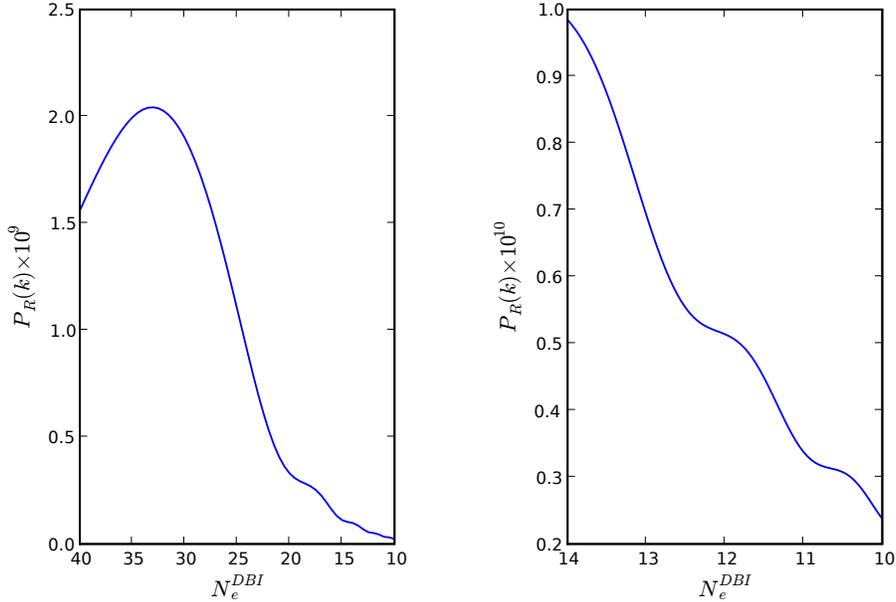}
\end{center}
\caption{\small Illustration of the effect of the non-BD vacuum and
the stringy phase transition in IR DBI model,
combining (\ref{PkallDBI}) and (\ref{PknonBD}). We use 
$N_c \sim 40$, which
is also the scale where $M\sim H/c_s$. The ripples at small
scales are due to
the non-BD effects and the suppression in large scales 
is due to the phase transition. The
connection between them is smoothed out by hand due to a lack of more
detailed understanding. The right panel is a blow-up of the left.}
\label{Fig:nonBD}
\end{figure}

We next look at the effect of the non-BD vacuum on the IR DBI
inflation. Now $\epsilon$ is negligibly small and 
$s=1/N_e^{\rm DBI}$. 
The natural scale of
$M$ is the red-shifted string scale, so $M \propto 1/N_e^{\rm DBI}$
and $\mu = 1/N_e^{\rm DBI}$. As we see from Sec.~\ref{Sec:Attractor},
$N_e^{\rm DBI}$ is typically smaller than $N_e$. 
So the variation of $n_s$ can be comparable
to $\CO(0.1)$. Using (\ref{DeltaA}), over $\CO(10)$ $e$-folds the change
of
the arguments in the trigonometric function is $2Mc_s/H \gg 1$, so the
modulation is oscillating rapidly. In fact since the larger scales are
associated with smaller $M$ during inflation, 
by the time that the modulation stops oscillating, we
have $M \sim H/c_s$. This is already 
beyond the validity region of the non-BD vacuum
calculations, and in fact is the place where the
stringy phase transition takes effect.
So we conclude that, in IR DBI model, 
the effect of the non-BD vacuum
is smoothly connected to the phase transition. 
Phenomenologically they look different from the
phase transition by causing frequent oscillations in the spectral
index. In Fig.~\ref{Fig:nonBD} 
we illustrate this effect on the power spectrum.
The amplitude of the oscillatory modulation increases and the
frequency decreases towards
the large scales, and finally merges into the phase transition. 

The effect of the non-BD vacuum on the large
non-Gaussianities in DBI inflation is studied in
\cite{Chen:2006nt}. The distinctive signature is the rising behavior
in the shape of the 3-point function in the folded triangle
limit. This
may also be an interesting clue to a better understanding of the
properties of the non-Gaussianity during the phase transition.

\section{Details of numerical calculations}
\label{Sec:Numerical}
\setcounter{equation}{0}

The zero mode motion of the brane in a warped throat is captured by
the DBI-CS action (\ref{DBICS}). Varying the action, the exact
form of the equations of motion is given by, 
\begin{eqnarray}
\phi_{NN} &=& -\frac{3}{2} \frac{f'}{f} \phi_N^2 - \left
( \frac{H_N}{H} + \frac{3}{\gamma^2}\right) \phi_N + \frac{f'}{f^2H^2}
- \frac{1}{\gamma^3H^2}\left(V' + \frac{f'}{f^2} \right) ~,
\label{ddphi}\\ 
H_N &=& -\frac{1}{2}\gamma H \phi_N^2 \label{dH} ~, \\
\gamma &=& \left[ 1 - f(\phi)H^2\phi_N^2  \right]^{-1/2} ~,\\
f(\phi) &=& \frac{\lambda}{\phi^4}, \quad \lambda \equiv n_BT_3R_B^4
~.
\end{eqnarray}
In the above differential equations, we choose to use the number of
$e$-folds $\tilde N_e = \ln a(t)$ as the time coordinate, as the intrinsic time
scale of inflationary dynamics is the Hubble time $H^{-1}$.  The
subscript $_N$ denotes derivatives with respect to $\tilde N_e$,
{\it{i.e.}}~$\phi_N \equiv d\phi / d\tilde N_e$, $\phi_{NN} \equiv d^2
\phi/d\tilde N_e^2$. We also denote derivatives with respect to $\phi$ by
a prime, {\it{i.e.}}~$f' \equiv df/d\phi$, $V' \equiv dV/d\phi$.

The equations (\ref{ddphi}) and (\ref{dH}) can be integrated
numerically using the conventional Runge-Kutta method. We note that
ignoring Eq.~(\ref{dH}) and setting $H_N/H = 0$ in Eq.~(\ref{ddphi})
does not introduce detectable errors to the results of
numerical calculation, because $H$ can safely be treated as a constant
for IR DBI inflation. Nevertheless, we have put Eq.~(\ref{dH}) through
numerical integration together with (\ref{ddphi})
for self-consistency.

To integrate the equation of motion, the code needs to know the
initial values $\phi(0)$ and $\phi_N(0)$. Due to the attractor
behavior of the IR DBI dynamics, these initial conditions 
will be irrelevant to our calculation of observables as
long as inflation lasts a few $e$-folds more than the minimum number
required to solve the horizon problem. This can always be done in the
IR DBI model, since we have the freedom to extend the start of
inflation to the IR end of the throat by choosing $\phi(0)$
appropriately. In practice, we choose $\phi(0) <
H\sqrt{\lambda_B}/80$, so that according to (\ref{NeDBI}), we will
have at least 80 DBI $e$-folds, and roughly 90 total $e$-folds
(assuming 10 non-relativistic $e$-folds). This is good enough to make
sure that, when we calculate primordial density perturbations on
scales relevant to CMB temperature anisotropies (roughly $50 \sim 60$
$e$-folds before the end of inflation), the inflationary background is
well on the attractor solution.  

To set up the model, the numerical code needs the five microscopic parameters 
$\{N_{B}$, $n_B$, $n_Ah_A^4$, $g_s^{-1/4}m_s$, $\beta\}$. The five input
parameters need to satisfy various bounds for model building
consistency. The following microphysical bounds are imposed in the
code:  
\begin{itemize}
\item the geometric constraint from compactification
(Eq.~(\ref{msbound}))
\[ \frac{m_s}{g_s^{1/4}} \lesssim 2^{3/2}\pi^{11/4} a_B^{1/2} \frac{\mpl}{N_B^{3/4}} ~;\]
\item the maximum number of branes generated by antibrane-flux
annihilation (Eq.~(\ref{nBbound}))
\[ n_B \lesssim \sqrt{N_B/(a_B g_s)}  ~;\]
\item the lower bound on string scale
\[ m_s g_s^{-1/4} \ge {\rm TeV} ~; \]
\item the lower bound on inflation scale
\[ n_A h_A^4 T_3 \ge {\rm TeV}^4 ~; \]
\item the warp factor $h_A\le 1$,
\[ n_A h_A^4 \le n_B ~.\]
\end{itemize}
(In these bounds, we fix the
string coupling $g_s = 0.1$ and $a_B =
1$. The effects of different $g_s$ and $a_B$ are discussed
in Sec.~\ref{Sec:Parameters}.)
Before calculating the density perturbation, the code performs checks
on all of the above bounds to make sure the input parameters are
theoretically consistent.

After numerically integrating the inflaton equation of motion, our
immediate result is $\phi(\tilde N_e)$, $\gamma(\tilde N_e)$ and
$H(\tilde N_e)$. Then we
use (\ref{Pkall}) to calculate the curvature perturbation $P_R(\tilde N_e)$
generated during inflation. Once we have $P_R(\tilde N_e)$, the horizon
crossing relation (\ref{Nek}) can translate $P_R(\tilde N_e)$ to
$P_R(k)$,
(noting $N_e= N_{\rm tot} - \tilde N_e$,)
which is then fed into CAMB to generate the CMB temperature anisotropy
spectrum and the matter power spectrum.

\end{document}